\documentclass{emulateapj}
\usepackage{color}

\bibpunct{(}{)}{;}{a}{}{,} 

\def\fm{\hbox{$.\!\!^{\rm m}$}}

\newcommand{\kms}{km\,${\rm s}^{-1}$\,}
\newcommand{\smy}{$[M_\odot\,{\rm yr}^{-1}]$\,}
\newcommand{\gcgs}{$[{\rm g}\,{\rm cm}^{-2}]$\,}
\newcommand{\dori}{\mbox{$\delta$\,Ori A}}

\shorttitle{A non-LTE analysis of $\delta$ Ori}

\begin{document}

\title{%
A coordinated X-ray and Optical Campaign of the Nearest Massive Eclipsing Binary, 
$\delta$~Orionis~A\lowercase{a}: 
IV. A multiwavelength, non-LTE spectroscopic analysis.}

\author{T. Shenar\altaffilmark{1}}
\author{L. Oskinova \altaffilmark{1}}
\author{W.-R. Hamann\altaffilmark{1}}
\author{M. F. Corcoran\altaffilmark{2,3}}
\author{A. F. J. Moffat\altaffilmark{4}}
\author{H. Pablo\altaffilmark{4}}
\author{N. D. Richardson\altaffilmark{4}}
\author{W. L. Waldron\altaffilmark{5}}
\author{D. P. Huenemoerder\altaffilmark{6}}
\author{J. Ma{\'\i}z Apell{\'a}niz\altaffilmark{7}}
\author{J. S. Nichols\altaffilmark{8}}
\author{H. Todt\altaffilmark{1}}
\author{Y. Naz\'e\altaffilmark{9,10}}
\author{J. L. Hoffman\altaffilmark{11}}
\author{A. M. T. Pollock\altaffilmark{12}}
\author{I. Negueruela\altaffilmark{13}}

\altaffiltext{1}{Institut f\"ur Physik und Astronomie, Universit\"at Potsdam, Karl-Liebknecht-Str. 24/25, D-14476 Potsdam, Germany}
\altaffiltext{2}{CRESST and X-ray Astrophysics Laboratory, NASA/Goddard Space Flight Center, Greenbelt, MD 20771}
\altaffiltext{3}{Universities Space Research Association, 10211 Wincopin Circle, Suite 500 Columbia, MD 21044, USA.}
\altaffiltext{4}{D\'epartement de physique and Centre de Recherche en Astrophysique 
du Qu\'ebec (CRAQ), Universit\'e de Montr\'eal, C.P. 6128, Succ.~Centre-Ville, Montr\'eal, Qu\'ebec, H3C 3J7, Canada}
\altaffiltext{5}{Eureka Scientific, Inc., 2452 Delmer St., Oakland, CA 94602}
\altaffiltext{6}{Kalvi Institute for Astrophysics and Space Research, MIT, Cambridge, MA, USA}
\altaffiltext{7}{Centro de Astrobiolog{\'\i}a, INTA-CSIC, Campus ESAC, P.O. Box 78, E-28\,691 Villanueva de la Ca\~nada, Madrid, Spain}
\altaffiltext{8}{Harvard-Smithsonian Center for Astrophysics, 60 Garden St., MS 34, Cambridge, MA 02138}
\altaffiltext{9}{FNRS Research Associate}
\altaffiltext{10}{Groupe d'Astrophysique des Hautes Energies, Institut d'Astrophysique 
et de G\'eophysique, Universit\'e de Li\'ege, 17, All\'ee du 6 Ao\^ut, B5c, B-4000 Sart Tilman, Belgium}
\altaffiltext{11}{Department of Physics and Astronomy, University of Denver,
  2112 E. Wesley Ave., Denver, CO, 80208 USA}
\altaffiltext{12}{European Space Agency, \textit{XMM-Newton} Science Operations Centre, European Space Astronomy Centre, 
Apartado 78, 28691 Villanueva de la Ca\~{n}ada, Spain}
\altaffiltext{13}{Departamento de F{\'\i}sica, Ingenier{\'\i}a de Sistemas y Teor{\'\i}a de la Se\~nal, 
Escuela Polit\'ecnica Superior, Universidad de Alicante, P.O.~Box~99, E-03\,080 Alicante, Spain}

\begin{abstract}
Eclipsing systems of massive stars 
allow one to explore the properties of their components in great detail.
We perform a multi-wavelength, non-LTE analysis of the three components of 
the massive multiple system $\delta$~Ori~A, focusing 
on the fundamental stellar properties, stellar winds, and X-ray characteristics of the system. 

The primary's distance-independent parameters turn out to be characteristic for its spectral 
type (O9.5~II), but usage of the {\it Hipparcos} parallax yields surprisingly low 
values for the mass, radius, and luminosity. Consistent values follow only if $\delta$ Ori  lies 
at about twice the {\it Hipparcos} distance, in the vicinity of the $\sigma$-Orionis cluster.  The primary and 
tertiary dominate the spectrum and leave the secondary only marginally 
detectable. We estimate the V-band magnitude difference between primary and secondary 
to be $\Delta V \approx 2\fm8$. The inferred parameters suggest the secondary is an early B-type dwarf 
($\approx$ B1~V), while the tertiary is an early B-type subgiant ($\approx$ B0~IV).
We find evidence for rapid 
turbulent velocities ($\sim 200$\,\kms) and wind inhomogeneities, partially optically 
thick, in the primary's wind. The bulk of the X-ray emission likely emerges from the 
primary's stellar wind ($\log L_{\text{X}} / L_{\text{Bol}} \approx -6.85$), initiating 
close to the stellar surface at $R_0 \sim 1.1\,R_*$. Accounting for clumping, the mass-loss rate of the primary 
is found to be $\log \dot{M} \approx -6.4\,$\smy, which agrees with hydrodynamic predictions,
and provides a consistent picture along the X-ray, UV, optical and radio spectral domains.
\end{abstract}


\section{Introduction}
\label{sec:introduction}

Massive stars ($M \gtrsim 10M_\odot$) bear a tremendous influence on their host 
galaxies, owing to their strong ionizing radiation and powerful stellar winds
\cite[e.g.,][]{Kudritzki2000, Hamann2006}. Yet our understanding of massive stars 
and their evolution still leaves much to be desired: (1) Values of mass-loss rates 
derived in different studies may disagree with each other by up to an order of 
magnitude \cite[e.g.,][]{Puls1996, Fullerton2006, Waldron2010, Bouret2012} 
and often do not agree with theoretically predicted mass-loss rates \cite[e.g.,][]{Vink2000}.
(2) The extent of wind inhomogeneities,
which greatly influence mass-loss rates inferred by means of spectral analyses, 
are still largely debated 
\cite[][]{Shaviv2000, Owocki2004, Oskinova2007, Sundqvist2011, Surlan2013}.
(3) The production mechanisms of X-ray radiation in massive stars
have been a central subject of study in recent decades \cite[][]{Feldmeier1997, 
Pollock2007} and are still far from being understood. 
(4) The effect of magnetic fields 
and stellar rotation on massive stars 
\cite[e.g.,][]{Friend1984, Maheswaran2009, Oskinova2011, DeMink2013, Petit2014, Shenar2014}, 
e.g., through magnetic braking \cite[][]{Weber1967} or chemical mixing \cite[][]{Maeder1987},
are still being investigated. (5) Lastly, stellar multiplicity 
seems to play a fundamental role in the context of massive stars,
significantly affecting their evolution
\cite[e.g.,][]{Eldridge2013}.

Several studies in the past years
\cite[e.g.,][]{Mason2009, Maiz2010, Chini2012, Sana2013, Sota2014, Aldoretta2014} give direct evidence 
that at least half of the massive stars are found in multiple systems. 
Massive stars in close binary systems generally 
evolve differently from single stars. Such systems may experience significant tidal forces
\cite[][]{Zahn1975, Palate2013}, mass-transfer \cite[][]{Pols1991}, additional supernova 
kicks \cite[][]{Hurley2002}, and mutual irradiation effects \cite[][]{Howarth1997}. Given the 
large binary fraction, an understanding of these processes is critical to
properly model the evolution of massive stars.
Fortunately, binary systems have two main advantages over single stars: First, they 
offer us the opportunity to empirically determine stellar masses.
Second, eclipsing binary systems provide the unique opportunity to investigate 
different characteristics of stellar winds by taking advantage of 
occultation  \cite[e.g.,][]{Antokhin2011}.
It is therefore insightful to analyze 
eclipsing multiple systems of massive stars in our Galactic neighborhood, using 
adequate, state-of-the-art modeling tools.

\begin{figure}[!htb]
\centering
  \includegraphics[width=\columnwidth]{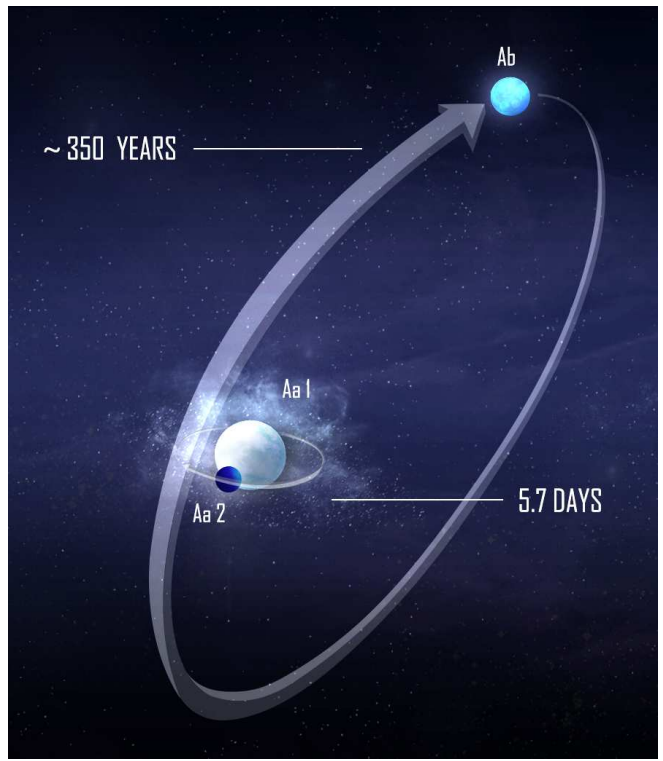}
  \caption{Artist's impression of the triple system $\delta$ Ori A, as viewed from Earth. The primary (Aa1) and Secondary (Aa2)
          form the tight eclipsing binary of period $5.7$\,d. The primary shows evidence for a significant wind 
          in all spectral domains. A third star (Ab) at an angular separation of $\approx 0.3\arcsec$ ($\sim 100\,$AU at 
          a distance of $d = 380$\,pc to Earth) orbits the system with a period
          of $\approx$346\,yr \citep{Tokovinin2014} and contributes roughly $25\%$ to the total visual flux.
          The sizes of all three components are drawn are to scale, as inferred in 
          this collaborative study. Their colors reflect their relative temperatures. Note that the tertiary is the second brightest
          companion in the system. The distance 
          between the binary system Aa and the tertiary Ab is not to scale. 
           Credits to Jessica Mayo}
\label{fig:isca}
\end{figure}

The star $\delta$ Ori A 
(HD~36\,486, Mintaka, HIP~25\,930, HR 1852)
is a massive triple system (see artist's illustration in Fig.\,\ref{fig:isca}) comprised of the close eclipsing binary
Aa (primary Aa1, secondary Aa2) with a 5.732\,d period
\cite[][]{Harvey1987, Harvin2002, Mayer2010},
and the more distant tertiary Ab at an angular separation of $\approx 0.3\arcsec$ 
relative to the binary Aa with a period
of $\sim$346\,yr \cite[][]{Heintz1980, Perryman1997, Tokovinin2014}.
The tertiary Ab has been photometrically resolved from the binary Aa in different 
surveys \cite[][]{Horch2001, Mason2009, Maiz2010}
and is found to contribute $\approx 25\%$  to the system's flux in the visual band.
These three components are not to be confused with the more distant and 
significantly fainter stars
$\delta$~Ori B and C at separation 33$\arcsec$ and 53$\arcsec$, respectively; 
together, these five stars comprise the multiple system $\delta$~Ori, also
known as Mintaka.
With a visual magnitude of $V = 2\fm24$ outside eclipse \cite[][]{Morel1978}, 
$\delta$~Ori~A is one of the brightest massive multiple systems in the 
night sky, making it easily visible to the naked eye (it is the westmost star in Orion's belt). 
According to the new {\it Hipparcos} 
reduction \cite[][]{VanLeeuwen2007}, its parallax is $4.71\pm0.58$\,mas, corresponding to
a distance of $d = 212\pm30$\,pc\footnote{This value is not corrected for the Lutz-Kelker effect \cite[][]{Lutz1973}. \cite{Maiz2008} account for this effect and revise
the distance to $d = 221$\,pc. However, the 
difference is negligible within measurement uncertainties.}.
On the other hand, the system resides in the Orion OB1b association, to which 
the $\sigma$-Orionis cluster belongs as well. The distance to the cluster itself 
is estimated to be $d \sim 380$\,pc, almost a factor 2 larger than the {\it Hipparcos} 
distance \cite[see][and references therein]{Caballero2008}.

This paper is a contribution to a series of papers 
within the framework of the $\delta$ Ori collaboration. The other 
papers include the analysis of high quality {\it Chandra} observations to explore 
X-ray properties (Corcoran et al.\ in prep.\ Paper I) and  variability 
(Nichols et al.\ in prep.\ Paper II) in the system, 
and a complete photometric and spectrometric variabilty analysis
(Pablo et al.\ in prep.\ Paper III). In this study, we focus on a non-LTE, 
multiwavelength spectral analysis of the three components Aa1, Aa2, and Ab,
with the goal of obtaining reliable stellar and wind parameters.

$\delta$~Ori~A has been repeatedly studied previously. 
The system, which shows clear evidence for a significant stellar wind in
the optical, UV, and X-ray domains, has been assigned the spectral type O9.5~II
\cite[][]{Walborn1972}, later refined by \cite{Sota2011} to O9.5~II~Nwk, which most 
probably corresponds to the brigtest component: Aa1.
With less confidence, the secondary Aa2 has been assigned the spectral type B0.5~III \cite[][]{Harvin2002}.
This result is questioned by \cite{Mayer2010}, who argue that the secondary 
is too faint for this spectral type. The latter authors 
further suggest the spectral type O9~IV for the tertiary, and leave the 
secondary unclassified.

Although the stellar parameters of an eclipsing stellar system
can usually be sharply constrained,  
much controversy is found in the literature in the case of $\delta$~Ori~A. \cite{Koch1987} reported 
$M_1 = 23\,M_\odot$, 
$R_1 = 17\,R_\odot$  and $M_2 = 9\,M_\odot$,
$R_2 = 10\,R_\odot$ for the primary and secondary masses and radii, respectively, 
as well as a V-band magnitude difference of 
$\Delta V_{\rm Aa1Aa2} = 1\fm4$. \cite{Harvin2002}
later inferred significantly smaller masses for the primary and secondary, $M_1 = 11.2\,M_\odot$, 
$M_2 = 5.6\,M_\odot$, 
and a smaller contribution of the secondary 
in the visual band: $\Delta V_{\rm Aa1Aa2} = 2\fm5$. These results
were challenged by 
\cite{Mayer2010}, who suggested that a confusion between the secondary  Aa2 and tertiary Ab 
led to the low masses obtained by \cite{Harvin2002}.
\cite{Mayer2010} inferred
$R_1 = 15.6\,R_\odot$ and $R_2 = 4.8\,R_\odot$. They did not detect
any contribution from the secondary Aa2 and concluded that $\Delta V_{\rm
  Aa1Aa2} \gtrsim 3\fm5$. Assuming $M_1 = 25\,M_\odot$, they
inferred $M_2 = 9.9\,M_\odot$. 
In Paper~III, the secondary's radial velocity (RV) curve could not be constructed,
and, for an adopted primary mass of 
$M_1 \approx 24\,M_\odot$, 
it was concluded that $M_2 = 8.5\,M_\odot$
$R_1 = 15.1\,R_\odot$, $R_2 = 5.0\,R_\odot$.

Similarly, reported values of the mass-loss rate of the 
primary are quite diverse:
\cite{Lamers1993} report
$\log \dot{M}_1 = -5.97\,$\,\smy based on radio observation and 
$\log \dot{M}_1 = -5.92\,$\,\smy based on H$\alpha$ analysis.  
\cite{Lamers1999} later similarly obtain $\log \dot{M}_1 = -5.7\,$\,\smy 
based on P Cygni profile analysis using the 
Sobolev plus exact integration (SEI) method. 
Neglecting the effects of wind inhomogeneities and porosity \cite[][]{Oskinova2007, Surlan2013}
on the observed X-ray spectrum, and adopting a generic O-type model,
\cite{Cohen2014} derived a value of $\dot{M}_1 = -7.2\,$\,\smy
from X-ray line profile fitting, more than an order of magnitude less 
than the previously inferred value.
Indeed, no consensus on the mass-loss rate of the primary has been reached so far.

$\delta$~Ori~A has an observed X-ray luminosity of \mbox{$L_{\rm X}\approx1.4\cdot10^{32}$\,erg\,${\rm s}^{-1}$} 
for $d = 380$\,pc (Paper I). $\delta$ Ori A's X-ray properties 
were previously explored by
\cite{Miller2002}, \cite{Leutenegger2006}, and \cite{Raassen2013}. These studies
generally identify the X-ray formation process to be intrinsic to the primary's wind, 
a result which is further supported within our collaboration.
An extensive review of past X-ray observations and analyses
are given by Papers I and II.

In this study, we perform a consistent non-LTE photosphere and wind analysis of
the three components of the triple system $\delta$~Ori~A in the optical, UV, and X-ray domains, 
at several orbital phases. 
We analyze the optical and UV spectra using 
the non-LTE Potsdam Wolf-Rayet (PoWR) code, which is applicable to any hot star. 
We further illustrate the importance of optically thin and 
optically thick clumps in the wind. 
We use the non-LTE models to simulate the effect of 
X-rays in the wind of the primary and derive onset radii of X-ray formation regions using 
ratios of forbidden and intercombination lines in the {\it Chandra} spectra. 
Finally, using the non-LTE model of the primary, 
we calculate synthetic X-ray line profiles and compare them to observed ones.
Our results are further compared to studies of the radio 
emission from the system. This study thus encompasses 
the whole range from the X-ray domain to the radio domain.

The structure of the paper is as follows: In Sect.~\ref{sec:obsdata} we 
describe the observational data used. Our modeling methods and assumptions are 
discussed in Sect.~\ref{sec:analysis}. In Sect.~\ref{sec:results}, we present and discuss our results.  
Sect.~\ref{sec:macroclump} focuses on the effect of wind inhomogeneities on the optical, UV and X-ray spectral domains, 
while in Sect.~\ref{sec:xrays} we study the X-ray radiation of the star. 
Lastly, in Sect.~\ref{sec:summary}, we summarize our results.

\section{Observational data}
\label{sec:obsdata}

All spectra used in this analysis contain the  contribution of the
primary Aa1, secondary Aa2, and tertiary Ab. In the following, 
all phases given are photometric phases relative to primary minimum (occurring 
when the secondary occults the primary), calculated with $E_0 = {\rm HJD}~2456277.79$\,d
and $P = 5.732436\,$d (Paper~III, and references therein).

Two of our optical spectra are of high resolution and high signal-to-noise, obtained with the 
NARVAL spectropolarimeter on 23-24 October
2008. These spectra were reduced with standard techniques and downloaded from the 
PolarBase\footnote{http://polarbase.irap.omp.eu} 
\cite[][]{Petit2014}. The observations were carried out with the Telescope Bernard Lyot. 
The spectra have a S/N of $500 - 800$, 
and correspond to
phases $\phi = 0.84$ and $\phi = 0.02$.

Three additional optical spectra at phases \mbox{$\phi$ = 0.19, 0.38, and 0.54} were obtained on the nights 
of 28, 29, and 30 Dec 2012 (contemporaneous with out Chandra and MOST observations) using the CAF\'E 
spectrograph at the 2.2 m Calar Alto telescope as part of the CAF\'E-BEANS
project, a survey that is obtaining high-resolution 
($R\sim 65\,000$) multi-epoch optical spectroscopy of all bright O stars in the 
northern hemisphere to complement OWN, the equivalent southern hemisphere survey 
\cite[][]{Barba2010, Sota2014}. On each date, ten consecutive 30\,s exposures
were obtained and combined to yield  spectra with S/N varying from $\sim200$ in the blue part
($\sim4500\,$\AA) to $\sim500$ in the red part ($\sim6000\,$\AA). The data were processed using 
a pipeline developed specifically for the project. The velocity stability was checked 
using ISM lines. More details regarding the reduction pipeline are given by 
\citet{Negueruela2014}.

For the spectral range $1200 - 2000$\,\AA, we make use of archival {\it IUE} spectra 
at different orbital phases (see the observation log given by \citealt{Harvin2002}). 
The spectra have a signal-to-noise (S/N) ratio of $\sim 10 /$pixel, and are averaged in bins of 
$0.1$\,\AA. The {\it IUE} spectra are flux-calibrated and are rectified using the model continuum.

In the spectral range $1000 - 1200$\,\AA, we make use of the 
{\it Copernicus} observation available in the {\it Copernicus}
archive under the ID c025-001.u2. This observation consists of 48 
co-added scans obtained between 21 and 24 November 1972.

For the spectral energy distribution (SED), we make use of 
$U$, $B$, $V$, $R$, $I$, $J$, $H$, $K$ \cite[][]{Morel1978}, 
WISE \cite[][]{Cutri2012}, and IRAS \cite[][]{Moshir1990} photometry. We further
use a low-resolution, flux calibrated optical spectrum kindly supplied to us 
by S.\ Fabrika and A.\ Valeev (priv.\ com.).
The spectrum was obtained with the Russian BTA telescope using the SCORPIO focal reducer, 
on 31 Dec 2013 in the range $3800 - 7200$\,\AA\ with a spectral resolution of $6.3$\,\AA, and corresponds
to phase $\phi = 0.42$. 
A Hartmann mask was used to avoid saturation.

The {\it Chandra} X-ray spectra used in Sect.\,\ref{sec:xrays} were taken with the HEG and 
MEG detectors for a total exposure time of 487.7\,ks. The data are
thoroughly described in Papers I and II.

\section{Non-LTE photosphere and wind modeling}
\label{sec:analysis}
\subsection{The PoWR code}
\label{subsec:code}

PoWR is a non-LTE model atmosphere
code especially suitable for hot stars with expanding atmospheres\footnote{PoWR models of Wolf-Rayet stars can be downloaded at
http://www.astro.physik.uni-potsdam.de/PoWR.html}. 
The code consists of two main parts. In the first part, referred to as
the non-LTE iteration, the co-moving frame
radiative transfer in spherical symmetry and the statistical balance equations  are solved in an iterative scheme under the constraint
of energy conservation, yielding
the occupation numbers in the photosphere and wind. The second
part, referred to as the formal integration, 
delivers a synthetic spectrum in the observer's
frame. The
pre-specified wind velocity field takes the form of a $\beta$-law \cite[][]{CAK1975}
in the supersonic region. In the subsonic region, the velocity
field is defined so that a hydrostatic density stratification is
approached. 
Line blanketing by the iron lines is treated in the
superlevel approach \cite[][]{Graefener2002}, as originally
introduced by \cite{Anderson1989}. 
A closer description of the assumptions and methods used
in the code is given by \cite{Graefener2002} and \cite{Hamann2004}.

In the non-LTE iteration, line profiles are Gaussians with a constant Doppler width $v_{\rm Dop}$. 
In the formal integration, the
Doppler velocity is composed of depth-dependent thermal motion and microturbulence. 
The microturbulence $\xi(r)$ is 
interpolated between the photospheric microturbulence $\xi(r_{\rm ph}) = \xi_{\rm ph}$ and 
the wind microturbulence $\xi(r_{\rm w}) = \xi_{\rm w}$, with the radii $r_{\rm ph}$ and
$r_{\rm w}$ pre-specified.
Thermal and pressure broadening are accounted for in the 
formal integration. 
Turbulence pressure is also accounted for in the non-LTE iteration. 
Optically thin 
wind inhomogeneities are accounted for in the non-LTE iteration using the so-called `microclumping'
approach \cite[][]{Hamann1995a}. The density
contrast between a clumped and a smooth wind with an identical
mass-loss rate is described by the depth-dependent clumping factor $D(r)$ \cite[][]{Hamann1998},
where the clumps are assumed to be optically thin. 
Optically thick clumps, or `macroclumps`, are accounted for in the formal integration 
as described by \cite{Oskinova2007}, where the
clump separation parameter $L_{\rm mac}$ is to be specified (see Sect.~\ref{sec:macroclump}).

Four fundamental input parameters define a model atmosphere of an OB type star: 
$T_\ast$, $L$, $g_\ast$, $\dot{M}$. 
$T_\ast$ is the effective temperature of a star
with luminosity $L$ and radius $R_\ast$, as defined by the
Stefan-Boltzmann relation $L=4\pi \sigma R_\ast^2 T_\ast^4$. 
$g_*$ is related to the radius $R_*$ and mass $M_*$  via $g_* = g(R_*) = G\,M_* R_*^{-2}$. 
$\dot{M}$ is the mass-loss rate of the star. The
stellar radius $R_\ast$ is defined at the continuum Rosseland optical depth
$\tau_\mathrm{Ross}$=20, which is the inner boundary of our model, and which was tested to be
sufficiently large.  
The outer boundary of the
model is set to $100\,R_*$.
Note that the stellar radius is generally not 
identical to the photospheric radius $R_{2/3}$ defined at
$\tau_\mathrm{Ross} = 2/3$. However, usually $R_* 
\cong R_{2/3}$, except in cases of 
extreme radiation pressures (e.g., supergiants, Wolf-Rayet stars). 
Nevertheless, one must bear in mind that the effective temperature
referring to the photospheric radius, which we denote 
with $T_{2/3}$ to avoid ambiguity, may slightly differ from $T_*$. 
Similarly, $g_{2/3} = g(R_{2/3})$ is the gravity at $\tau_\mathrm{Ross} = 2/3$.
Like most studies, we specify photospheric values when compiling our results in 
Table\,\ref{tab:stellarpar},
but the cautious reader should be
aware of this difference when comparing with other studies.

Due to the strong radiative pressures in massive stars, one cannot measure the 
gravity $g_*$ 
directly from their spectra, but rather the effective gravity $g_{\rm eff}$. 
$g_*$ is obtained from $g_{\rm eff}$ via
$g_{\rm eff} := g_* \left(1 - \overline{\Gamma}_{\rm rad}\right)$,
where $\overline{\Gamma}_{\rm rad}$ is a weighted average of the ratio of total outward radiative force to inward gravitational 
force over the hydrostatic domain. The outward radiative force is calculated consistently in the non-LTE iteration, 
and includes the contribution from line, continuum, and Thomson opacities (Sander et al.\ submitted).

\subsection{The analysis method}
\label{subsec:method}

The analysis of stellar spectra with non-LTE model atmosphere codes is
an iterative, computationally expensive process, which involves 
a multitude of parameters. Nevertheless, 
most parameters affect the spectrum uniquely.
Generally, the gravity $g_*$ 
is inferred from the wings of prominent 
hydrogen lines. 
The stellar temperature $T_*$ is obtained from the line ratios  
of ions belonging to the same element. 
Wind parameters such as $\dot{M}$, $v_\infty$, and $L_{\rm mac}$ are
derived from emission lines, mainly in the UV.
The luminosity $L$ and the colour excess $E(B-V)$ are determined 
from the observed spectral energy distribution (SED), and  
from flux-calibrated spectra. The abundances are determined from the overall
strengths of lines belonging to the respective elements. Finally, parameters describing 
the various velocity fields in the photosphere and wind (rotation, turbulence) 
are constrained from profile shapes and strengths. The radius $R_*$ and 
spectroscopic mass $M_*$ follow from $L$, $T_*$ and $g_*$.

To analyze a multiple system such as $\delta$~Ori~A, a model for each component star is required. 
As opposed to single stars, the luminosities of the components 
influence their contribution to the overall flux and thus affect the 
normalized spectrum.
The light ratios of the different components therefore become entangled with the 
fundamental stellar parameters.
Fortunately, with fixed temperatures and gravities for the secondary and 
tertiary components, 
the observational constraints
provide us with the light ratios (see Sect.~\ref{subsec:assumptions}).

Methods to disentangle a composite spectrum 
into its constituent spectral components by observing the system at different phases
have been proposed and implemented during the past few years \cite[][]{Bagnuolo1991, Simon1994, Hadrava1995, 
Gonzalez2006, Torres2011}.  
In fact, 
an attempt to disentangle the spectrum of $\delta$ Ori A was 
pursued by \cite{Harvin2002} 
and \cite{Mayer2010}. However, 
even after performing the disentanglement, 
the two sets of authors come to significantly different results.
A disentanglement of the He\,{\sc i} $\lambda 6878$ 
line was performed in Paper III, but, after accounting for the contribution of the tertiary as obtained here, 
no clear signal from the secondary was detected.
Given the very low contribution of the secondary Aa2, and having
only poor phase coverage in the optical, 
we do not pursue a disentanglement of $\delta$~Ori~A. 

As a first step, the secondary and tertiary models are
kept fixed. Motivated by the results of Paper~III, we initially adopt  $T_{2/3} \sim 25$\,kK and 
$\log g_{2/3} \sim 4.15$\,\gcgs for the secondary. The tertiary 
is initially fixed with the parameters suggested
by \cite{Mayer2010}. With the light ratios at hand, this fixes the luminosities of Aa2 and Ab.
Having fixed the secondary and tertiary models, 
we turn to the second step, which is an accurate analysis of the primary.

To constrain $T_*$, we use mostly He\,{\sc i} and He\,{\sc ii} lines, 
such as He\,{\sc i}  $\lambda\lambda$4026, 4144, 4388, 4713, 4922, 5015, 6678 and  
He\,{\sc ii} $\lambda$4200, 4542, 5412, 6683. The prominent 
He\,{\sc ii} $\lambda4686$ line is found to be a poor temperature indicator
(see Sect.~\ref{subsec:contribution}). 
The temperature is further 
verified from lines of carbon, nitrogen and silicon. 
$g_*$ is primarily derived from the wings
of prominent Balmer and He\,{\sc ii} lines. 
Here we encounter the difficult problem of identifying 
the contribution of the tertiary to the hydrogen lines due to the pronounced
wings of the Balmer lines. We therefore also made use of diagnostic lines such 
as C\,{\sc iii} $\lambda5696$, whose behavior heavily depends on the gravity 
(see Sect.\,\ref{subsec:errors}). 

\begin{table*}[!htb]
\caption{Inferred stellar parameters for the multiple stellar system $\delta$ Ori A} 
\label{tab:stellarpar}
\begin{center}
\begin{tabular}{lcccccc}
\hline
 & \multicolumn{2}{c}{Aa1} & \multicolumn{2}{c}{Aa2} & \multicolumn{2}{c}{Ab} \\ 
\hline
$T_{2/3}$ [kK] &  \multicolumn{2}{c}{29.5$\pm 0.5$} &   \multicolumn{2}{c}{25.6$\pm3000$} &  \multicolumn{2}{c}{28.4$\pm1500$}   \\
$\log g_{\rm eff}$ [cm\,${\rm s}^{-2}$] &  \multicolumn{2}{c}{3.0$\pm0.15$} &  \multicolumn{2}{c}{3.7} &  \multicolumn{2}{c}{3.2$\pm0.3$}  \\
$\log g_{2/3}$ [cm\,s$^{-2}$] & \multicolumn{2}{c}{3.37$\pm 0.15$} & \multicolumn{2}{c}{3.9} & \multicolumn{2}{c}{3.5$\pm0.3$} \\
$v_{\infty}$ [\kms] &  \multicolumn{2}{c}{2000$\pm100$}  &  \multicolumn{2}{c}{1200} &  \multicolumn{2}{c}{2000} \\
$D$ &  \multicolumn{2}{c}{10} &  \multicolumn{2}{c}{10} &  \multicolumn{2}{c}{10} \\
$L_{\rm mac}$ [$R_\odot$] &  \multicolumn{2}{c}{$\gtrsim$0.5}  &  \multicolumn{2}{c}{-} &  \multicolumn{2}{c}{-} \\
$E(B-V)$ [mag] &  \multicolumn{2}{c}{0.065$\pm0.01$} &  \multicolumn{2}{c}{0.065$\pm0.01$} &  \multicolumn{2}{c}{0.065$\pm0.01$}  \\
$A_V$ [mag] & \multicolumn{2}{c}{0.201$\pm0.03$} & \multicolumn{2}{c}{0.201$\pm0.03$} & \multicolumn{2}{c}{0.201$\pm0.03$} \\
$v \sin i$ [\kms] &  \multicolumn{2}{c}{130$\pm10$} &  \multicolumn{2}{c}{150$\pm50$} &  \multicolumn{2}{c}{220$\pm20$} \\
$\xi_{\rm ph}$ [\kms] &  \multicolumn{2}{c}{20$\pm5$} &  \multicolumn{2}{c}{10} &  \multicolumn{2}{c}{10$\pm5$}   \\
$\xi_{\rm w}$ [\kms] &  \multicolumn{2}{c}{200$\pm100$} &  \multicolumn{2}{c}{10} &  \multicolumn{2}{c}{10}   \\
$v_{\rm mac}$ [\kms] &  \multicolumn{2}{c}{60$\pm30$} &  \multicolumn{2}{c}{0} &  \multicolumn{2}{c}{50$\pm30$}  \\
H (mass fraction) &  \multicolumn{2}{c}{0.70$\pm0.05$} &  \multicolumn{2}{c}{0.7} &  \multicolumn{2}{c}{0.7} \\
He (mass fraction) &  \multicolumn{2}{c}{0.29$\pm0.05$} &  \multicolumn{2}{c}{0.29} &  \multicolumn{2}{c}{0.29} \\
C (mass fraction) &  \multicolumn{2}{c}{$2.4\pm 1\times10^{-3}$} &  \multicolumn{2}{c}{$2.4\times10^{-3}$} &  \multicolumn{2}{c}{$2.4\times10^{-3}$} \\
N (mass fraction) &  \multicolumn{2}{c}{$4.0\pm 2\times10^{-4}$} &  \multicolumn{2}{c}{$7.0\times10^{-4}$} &  \multicolumn{2}{c}{$7.0\times10^{-4}$} \\
O (mass fraction) &  \multicolumn{2}{c}{$6.0\pm 2\times10^{-3}$} &  \multicolumn{2}{c}{$6.0\times10^{-3}$} &  \multicolumn{2}{c}{$6.0\times10^{-3}$} \\
Mg (mass fraction) &  \multicolumn{2}{c}{$6.4\times10^{-4}$} &  \multicolumn{2}{c}{$6.4\times10^{-4}$} &  \multicolumn{2}{c}{$6.4\times10^{-4}$} \\
Al (mass fraction) &  \multicolumn{2}{c}{$5.6\times10^{-5}$} &  \multicolumn{2}{c}{$5.6\times10^{-5}$} &  \multicolumn{2}{c}{$5.6\times10^{-5}$} \\
Si (mass fraction) &  \multicolumn{2}{c}{$4\pm2\times10^{-4}$} &  \multicolumn{2}{c}{$6.6\times10^{-4}$} &  \multicolumn{2}{c}{$6.6\times10^{-4}$} \\
P (mass fraction) &  \multicolumn{2}{c}{$5.8\times10^{-6}$} &  \multicolumn{2}{c}{$5.8\times10^{-6}$} &  \multicolumn{2}{c}{$5.8\times10^{-6}$} \\
S (mass fraction) &  \multicolumn{2}{c}{$3.0\times10^{-4}$} &  \multicolumn{2}{c}{$3.0\times10^{-4}$} &  \multicolumn{2}{c}{$3.0\times10^{-4}$} \\
Fe (mass fraction) &  \multicolumn{2}{c}{$1.3\times10^{-3}$} &  \multicolumn{2}{c}{$1.3\times10^{-3}$} &  \multicolumn{2}{c}{$1.3\times10^{-3}$}  \\
\hline
Adopted distance $d$ [pc] & 212 & 380 & 212 & 380 & 212 & 380 \\ 
\hline
$\log L$ [$L_\odot$] & 4.77$\pm0.05$ & 5.28$\pm0.05$ & 3.7$\pm0.2$ & 4.2$\pm0.2$ & 4.3$\pm0.15$ & 4.8$\pm0.15$ \\
$\log \dot{M}$ [$M_\odot / {\rm yr}$] & $-$6.8$\pm0.15$ & $-$6.4$\pm0.15$ & $-$8.5 & $-$8.1 & $\le-$7.0 & $\le-$6.6   \\
$M_{\rm spec}$ [$M_\odot$] & 7.5$^{+3}_{-2.5}$ & 24$^{+10}_{-8}$ & 2.6 & 8.4  & 7.0$^{+7}_{-4}$ & 22.5$^{+24}_{-14}$ \\
$R_{2/3}$ [$R_\odot$] & 9.2$\pm0.5$ & 16.5$\pm1$ & 3.6$\pm1$ & 6.5$^{+2}_{-1.5}$& 5.8$\pm1$ & 10.4$\pm2$ \\
$M_V$ [mag] & $-$4.47$\pm0.13$ & $-$5.74$\pm0.13$ & $-$1.7$\pm0.5$ & $-$3.0$\pm0.5$ & -3.2$\pm0.4$ & -4.5$\pm0.4$ \\
\hline
\end{tabular}
\end{center}
\end{table*}

$\dot{M}$ and $L_{\rm mac}$
follow from a simultaneous fitting of H$\alpha$ and 
UV P Cygni lines. The wind parameters are checked for consistency
with previous analyses of radio observations and with X-ray observations 
(see Sect.\,\ref{subsec:xraysmod}).
The terminal velocity $v_\infty$ is determined from UV resonance lines. 
However, $v_\infty$ can only be determined accurately after the wind 
microturbulence has been deduced.
The determination of the projected rotational velocity $v \sin i$ and  
of the 
inner (photospheric) and outer (wind) 
microturbulent velocities $\xi_{\rm ph}$ and $\xi_{\rm w}$, 
as well as of the
macroturbulent velocity $v_{\rm mac}$, is discussed in detail
in Sect.~\ref{subsec:motions}. 

The abundances are determined from the overall strengths of
lines belonging to the respective elements. We include the elements 
H, He, C, N, O, Mg, Al, Si, P, S, and elements belonging to the iron group (e.g., Fe, Ni, Cr etc.). 
With the remaining stellar parameters fixed, 
this is  a straight-forward procedure. 

The total bolometric luminosity $L = L_1 + L_2 + L_3$ is obtained by fitting
the synthetic flux with the observed SED. The individual component luminosities
follow from the light ratios.
The color excess $E_{B - V}$ is derived using extinction laws published by \cite{Cardelli1989} with $R_V = 3.1$, and
can be very accurately determined from flux-calibrated {\it IUE} 
observations. The inferred value for $A_V$ is consistent with those of Ma{\'{\i}}z Apell{\'a}niz \& Sota (in
prep.\ 2015), who obtain a value of 0.185$\pm$0.013 mag using Tycho+Johnson+2MASS photometry and the extinction laws of \cite{Maiz2014}.

After obtaining a satisfactory model for the primary, we continue to
iterate on the secondary and tertiary models by identifying small
deviations between the composite synthetic and observed spectra at different 
orbital phases, as described in
Sect.~\ref{subsec:contribution}.  This allows us to adjust 
the temperatures and gravities of both models, as well as of the
projected rotation velocities. After improving the secondary 
and tertiary models, we return to the primary and adjust the model 
parameters accordingly. We repeat this process 
several times until a satisfactory fit is obtained in the UV and
optical, taking hundreds of lines into account in this process.

\subsection{Initial assumptions}
\label{subsec:assumptions}

Given the large number of parameters involved in this analysis, 
it is advisable to initially fix parameters which are constrained based on
observations, 
previous studies, and theoretical predictions. 

We adopt $\beta$=0.8 for the exponent
of the $\beta$-law for all components, which is both supported by observations as well as 
theoretically predicted \cite[e.g.,][]{Kudritzki1989, Puls1996}. 
Varying this parameter in the range 
$0.7 - 1.5$ 
did not significantly effect on the resulting synthetic spectrum.

A V-band magnitude difference of  $\Delta V_{\rm AaAb} = 1\fm35$ between the 
binary system Aa (composed of Aa1 and Aa2) and the tertiary Ab was 
measured by the {\it Hipparcos} satellite \cite[][]{Perryman1997}. \cite{Horch2001} report
$\Delta V_{\rm AaAb} = 1\fm59$, obtained from speckle photometry. \cite{Mason2009} find $\Delta V_{\rm AaAb} = 1\fm4$ by 
means of speckle interferometry. Like \cite{Mayer2010}, we adopt
$\Delta V_{\rm AaAb} = 1\fm4$, corresponding to 
the flux ratio $R_{AaAb} := F_V(Aa) /
F_V(Ab) = 3.63$. 
As for the binary components Aa1 and Aa2,
additional information regarding $\Delta V_{\rm Aa1Aa2}$ can be obtained from
the visual light curve of the system $\delta$~Ori~A. 
The
secondary light curve minimum (primary star in front) has a depth
$\Delta V_{\rm II. min} \approx 0\fm055$ with an error of about $0\fm005$.
Since the secondary eclipse is partial (Paper III), the secondary minimum
yields a lower limit to 
$R_{Aa2Aa1}$.
One obtains after some algebra

\begin{equation}
\label{eq:Aa2LR}
   R_{Aa2Aa1} \ge \frac{(R_{\rm II.min} -1)(R_{\rm AaAb} + 1)}{R_{\rm AaAb} - R_{\rm II.min} + 1}.
\end{equation}

In our case, $R_{\rm AaAb} \approx 3.63$ and $R_{\rm II.min} \approx 1.05$, and so 
$F_V(Aa2) \gtrsim 0.07\,F_V(Aa1)$,  which in turn implies \mbox{$\Delta V_{Aa1Aa2} \lesssim 2\fm9$}. It follows that 
the secondary Aa2 contributes \emph{at least} 
$6.5\%$ to the visual flux of the binary, and \emph{at least} $5.1\%$ to the total visual
flux of the system. We thus initially assume $\Delta V_{Aa2Aa1} = 2\fm8$, and further constrain 
this value  in the spectral \mbox{analysis} 
(see Section\,4.1).

In Sect.\,\ref{subsec:distance}, we show that usage of the {\it Hipparcos} distance $d = 212\,$pc results in
extremely peculiar parameters for the primary given its spectral 
type, and that a much better agreement is obtained for the alternative distance of the neighboring 
stellar cluster $\sigma$-Orionis, $d\sim380$\,pc \cite[see also discussion by][]{SimonDiaz2015}.
We therefore refer to both distances in the following, and discuss this issue more thoroughly 
in Sect.\,\ref{subsec:distance}. We note, however, that our results can be easily scaled with the distance, 
should it be revised in future studies (see Sect.\,\ref{sec:results}).

As we illustrate in Sect.~\ref{sec:macroclump}, there are several
indications for wind inhomogeneities (clumps) in the wind of Aa1. 
While clumping may already initiate at sub-photospheric layers 
e.g.,\ due to sub-photospheric convection \cite[][]{Cantiello2009}, we avoid an attempt 
to treat clumpiness in the optically thick layers below the photosphere.   
For reasons which are 
discussed in detail in Sect.\,\ref{sec:macroclump}, we adopt a depth-dependent clumping 
factor $D(r)$ which is fixed to 1 (no clumping) in the domain $R_* \le r \le 1.1\,R_*$ 
and reaches a maximum value of $D_{\rm max} = 10$ at $r \ge 10\,R_*$. 

Due to the faintness of the secondary and tertiary, we 
we can only give upper limits to their mass-loss rates, and set their terminal velocities
to be 2.6 times their escape velocity \cite[][]{Cassinelli}. We further assume 
$\xi(r) = 10$\,\kms\ for the secondary, a typical value for OB type stars of luminosity 
classes III and lower \cite[cf.][and references therein]{Villamariz2000}.
We choose to adopt an identical density contrast $D(r)$ for all three models.
Finally, we do not attempt to constrain the abundances in the secondary and
tertiary, but rather adopt solar values \cite[][]{Asplund2009}. This
set of assumptions should bear very little effect on the derived fundamental parameters 
of the three components.

\section{Results}
\label{sec:results}

The stellar parameters inferred
for the three components are given in Table\,\ref{tab:stellarpar}.
The model of the primary
includes the effect of macroclumping and X-rays, which we thoroughly discuss in Sects.\,\ref{sec:macroclump} and \ref{sec:xrays}. 
The upper part of the table displays parameters which, to first order, do not depend
on the adopted distance. 
The lower part of the table denotes distance-dependent parameters. For the convenience 
of the reader, we give these parameters for both ``candidate'' distances, $d = 212\,$pc and $d = 380\,$pc. 
Luminosities scale as $L \propto d^2$,  mass-loss rates as
$\dot{M} \propto d^{3/2}$, radii as $R_* \propto d$, and the mass as 
$M_{\rm spec}\propto d^2$.
The error margins given in Table~\ref{tab:stellarpar} are discussed in Sect.~\ref{subsec:errors} and 
are based on sensitivity of our analysis to these parameters.
Values without errors which are not upper bounds indicate adopted values.

\begin{figure*}[!htp]
\centering
  \includegraphics[width = .9\textwidth]{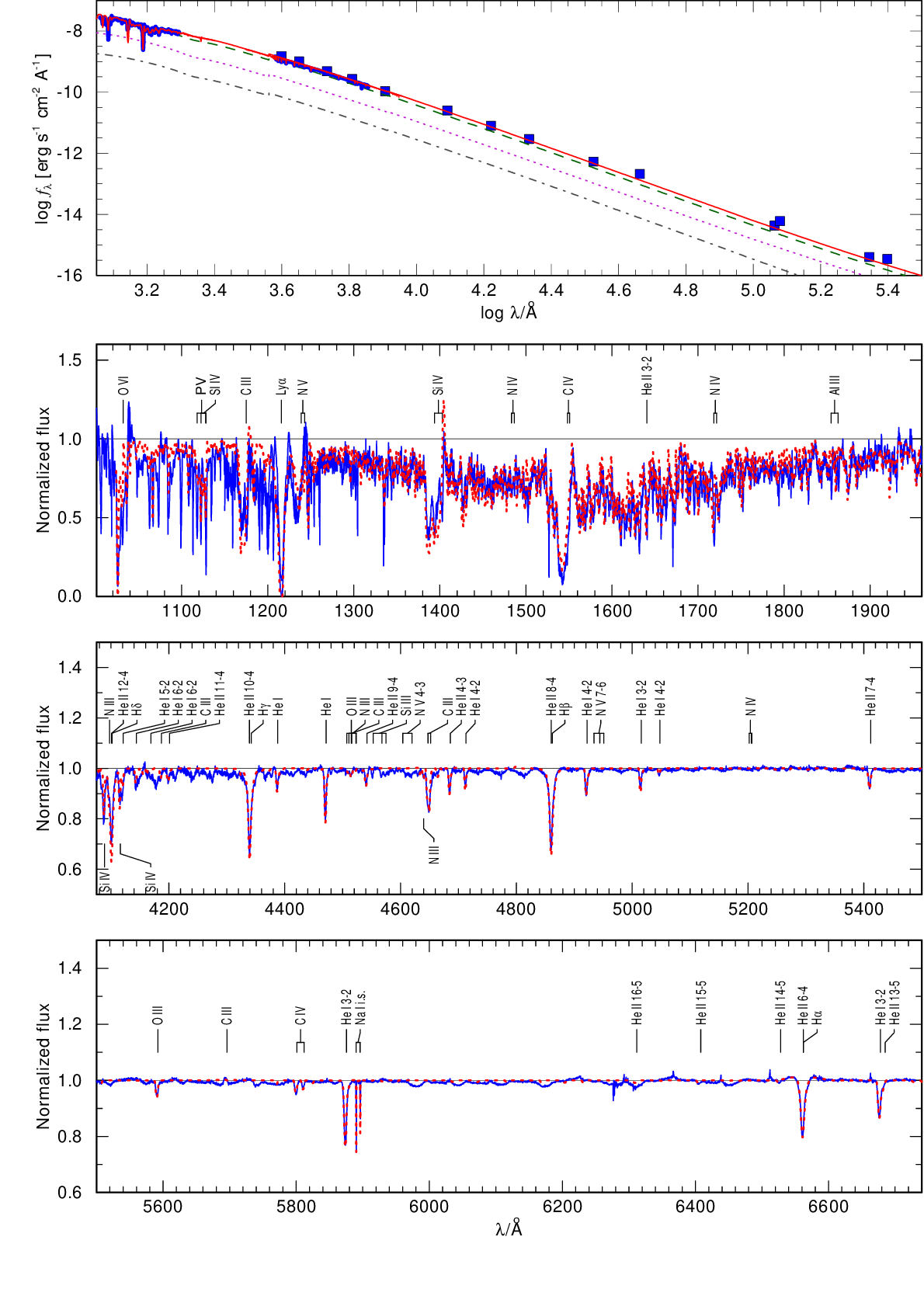}
  \caption{\textit{Upper panel:} Comparison between observations
    (blue squares and lines) and the total synthetic SED (red solid line), 
          which consists of the primary (green dashed line), secondary
          (gray dash-dotted line), and tertiary (pink dotted line)
          models.  
          \textit{Lower panels:} Comparison between the composite
          synthetic (red dotted line) and observed (blue line) normalized 
          spectra in the UV and optical. Both spectra roughly correspond to phase
          $\phi = 0.54$. For clarity, we refrain from showing the contributions of each
          component model to the normalized spectrum. The wavy pattern seen in the 
          observed spectrum (e.g.,\ in $\approx 6000 - 6200$\,\AA) is an artifact caused by connecting the echelle orders, 
          not related to the stellar system.}
\label{fig:tripfit}
\end{figure*}

The upper panel of Fig.\,\ref{fig:tripfit} displays the synthetic SEDs for the
models of the three components of the system.
The total synthetic SED (red solid line) comprises the
synthetic fluxes of the primary Aa1 (green dashed line), secondary Aa2 
(gray dotted-dashed line), and tertiary Ab (pink dotted line). Recall that 
the tertiary is brighter than the secondary. 
The blue squares denote $U$, $B$, $V$, $R$, $I$, $J$, $H$, $K$, 
WISE, and IRAS photometry. 
We also plot two observed flux-calibrated spectra in the UV and 
optical (blue lines). 
The lower panels of Fig.\,\ref{fig:tripfit} show the composite synthetic
(red dotted line) and observed (blue line) normalized UV 
and optical spectra of the system. The optical spectrum and the UV
spectrum in the range $1200 - 2000$\AA\ correspond to phase
$\phi \approx 0.54$. 
The synthetic normalized spectrum consists of three models calculated for  Aa1, Aa2, and Ab, 
shifted according to their RVs. 
The synthetic spectra are convolved with a Gaussian of FWHM = $0.1$\,\AA\ to 
account for instrumental broadening, inferred from fitting 
of the interstellar Na\,{\sc i} $\lambda\lambda 5890,\,5896.3$ lines.
We do not plot the 
individual spectra of the three components  for clarity.
Note that the wavy pattern seen in the 
observed spectrum (e.g.,\ in the domain $5900-6500$\,\AA) is an artifact which originates
in the connection of the echelle orders, not related to the stellar system.

It is evident that  both the
synthetic SED and normalized spectrum agree well with the observed spectrum. 
A good balance is obtained for all He\,{\sc i} lines and
He\,{\sc ii} lines, as well as for the metal lines. The inferred
parameters for microturbulence, rotation and macroturbulence yield
consistent line strengths and profiles over the whole spectral domain. The pseudo continuum
formed by the iron
forest in the range $1300 - 1800$\,\AA, as well as most photospheric and wind features, are
well reproduced.

The few features which are not reproduced very well are the Balmer lines,
and especially H$\delta$. H$\delta$ has a significantly 
smaller observed EW ($\approx 1.85$\,\AA) compared to the
synthetic spectrum ($\approx 2.35$\,\AA). In fact, the observed EW 
of H$\delta$ is somewhat smaller than typical for similar spectral types \cite[e.g.,][]{Cananzi1993},
and the question arises as to the cause. A significantly smaller gravity for the primary does 
not agree with the wing
shape of the other Balmer and He\,{\sc ii} lines and hardly affects the EW.  Reducing
the hydrogen abundance of the primary implies a larger helium abundance, 
which is not consistent with the helium lines. Larger $\dot{M}$ values 
lead to very strong emission in H$\alpha$, which is not
observed. The photospheric microturbulence has only negligible
effect on the EW of hydrogen lines. We therefore conclude that the
Balmer lines are dilluted by the light of one or both of the other components.
However, the low EW of H$\delta$
could only be reproduced
when assuming very peculiar parameters for the tertiary,
e.g., a very weak gravity ($\log g_{2/3} \lesssim 3.2\,$\gcgs), or very large mass-loss
rates ($\log \dot{M} \gtrsim -5.5\,\left[M_\odot / {\rm yr}\right]$). 
Such parameters are not only hard to justify physically, but also not
consistent with the remaining spectral lines.
The problem is also seen, albeit to a lesser extent, in the lines H$\gamma$ and H$\beta$. 
Future observations should shed light on this peculiarity.

\subsection{Which distance is the correct one?}
\label{subsec:distance}

It is reassuring that those fundamental stellar parameters of the primary which do not depend on the distance 
match well with its spectral type. Interpolating 
calibrations by \cite{Martins2005} to an O9.5~II class yields 
$T_{2/3} = 29.3$\,kK and $\log g_{2/3} = 3.35$\,\gcgs, which agrees   
with our results within the error margins.
The primary's nitrogen and silicon abundances are found to be 
slightly subsolar. It is interesting to note
that \cite{Sota2011} recently added the suffix ``Nwk'' to the spectral
type of Aa1, implying relatively weak nitrogen
lines, which we confirm  independently.

For $d = 380$\,pc, 
the distance-dependent parameters agree well with calibrations 
by \cite{Martins2005}.
Interpolating to luminosity class II, calibrations by \cite{Martins2005} imply
$\log L/L_\odot = 5.3$, $M = 25\,M_\odot$, $R_{2/3} =
18.2\,R_\odot$, and $M_V = -5\fm73$ for an O9.5~II star, which is consistent with 
our results. 
Furthermore, this distance implies radii and masses for the primary and secondary 
which agree well with the results obtained independently in Paper~III.
However, very peculiar values are obtained when using the {\it Hipparcos} distance
of $d = 212\,$pc (see lower part of Table\,\ref{tab:stellarpar}).
In fact, all three components appear to be peculiar
when adopting the {\it Hipparcos} distance. While a membership in a close binary
system offers some room
for deviations from standard values, this would not explain why the distant
tertiary Ab should be peculiar too. The fact that
distance-independent parameters 
are not unusual raises even more
suspicion regarding the {\it Hipparcos} distance.

A similar discrepancy is observed other bright stars: 
\citet{Hummel2013} analyzed the binary system 
$\zeta$~Orionis~A ($V = 1.79$) and inferred a distance of $d = 294\pm21\,$pc based on an orbital analysis, and 
$d = 387\pm54\,$pc based on a photometric estimate. 
Both these distances are significantly larger than the corresponding {\it Hipparcos} 
distance of $d = 225^{+38}_{-27}\,$pc \citep{VanLeeuwen2007}, which encouraged them to discard the {\it Hipparcos} distance in their study. 
Another bright star for which the {\it Hipparcos} parallax implies very peculiar stellar parameters is the prototypical O supergiant $\zeta$ Puppis ($V = 2.25$); 
Its distance 
was suggested to be at least twice as large as implied by its measured parallax \citep{Pauldrach2012, Howarth2014}.

Returning to $\delta$ Ori A, our estimate of $d \sim 380\,$pc would correspond to
a parallax of \mbox{$\pi \sim 2.6\,$mas}, deviating by $\sim3.5\sigma$ from the newly reduced {\it Hipparcos} parallax of $4.71\pm0.58$\,mas \citep{VanLeeuwen2007}.
The probability for such a deviation to arise randomly is thus extremely small ($<0.1\%$). Ground-based parallaxes for $\delta$ Ori A 
generally suffer from much larger uncertainties. For example, the Yale catalog gives $\pi = 9.7\pm6.7$\,mas \citep{Yale1995}.
It is interesting to note that the original {\it Hipparcos} catalog 
gives a parallax of $3.56\pm0.83\,$mas \citep{Hipcatalog1997}. In this case, 
the deviation could be plausibly explained as a random error, with $2.6\,$mas being 
$\sim 1\sigma$ away.
While stellar multiplicity has been 
suggested to cause systematic errors in parallax measurements \cite[e.g.][]{Szabados1997}, the argument is unlikely to hold here given the 
different timescale of the involved orbits compared to that of the parallax measurement (which is to say, a year). 
However, large errors in the parallax measurement are also expected to occur in the case of very bright stars, which could lead to saturation, resulting
in an inaccurate estimate of the barycenter of the point spread function.
Brightness is indeed a property $\delta$ Ori~A shares with the  
objects mentioned above. 
It is beyond the scope of the paper to judge whether the new {\it Hipparcos} reduction suffers 
from underestimated errors, but the fact that several bright stars show a similar pattern should encourage thorough studies on the matter.

The distance to $\delta$ Orionis is difficult to directly measure with modern techniques. 
The system is probably too bright for ground-based (e.g.\ RECONS\footnote{www.recons.org})
as well as space-based instruments (e.g.\ the {\it Hubble Space Telescope}'s Fine Guidance Sensor; the {\it Gaia} mission).  Our
best hope is to obtain the RV curve of the secondary, which requires optical spectroscopy with a S/N $\gtrsim1000$ (Paper III). Then, with
a direct measure of the system parameters, follow-up long-baseline interferometry may provide the angular extent of the orbit,
allowing for an orbital parallax measurement similar to that of $\zeta$ Ori A \citep{Hummel2013}. At 380\,pc, the angular extent
of the orbit will be on the order of 0.5\,mas and thus difficult to measure, as the smallest angular separation for a binary yet
resolved is 1.225\,mas \citep{Raghavan2009}. However, this was done in the $K-$ band with the CHARA Array, so the resolution
in the $R-$band should be sufficient to resolve the binary. 
Moreover, along with a light-curve analysis, and possessing \emph{both} RV curves, one could accurately derive the masses and radii 
of both components, which, upon comparison with a distance-dependent spectral analysis, would supply another constraint on the distance.
For now, we suggest that the {\it Hipparcos}
distance may be underestimated, but leave the question 
open for future studies to resolve.

%
%

\begin{figure*}[!htb]
\centering
  \includegraphics[width = \textwidth]{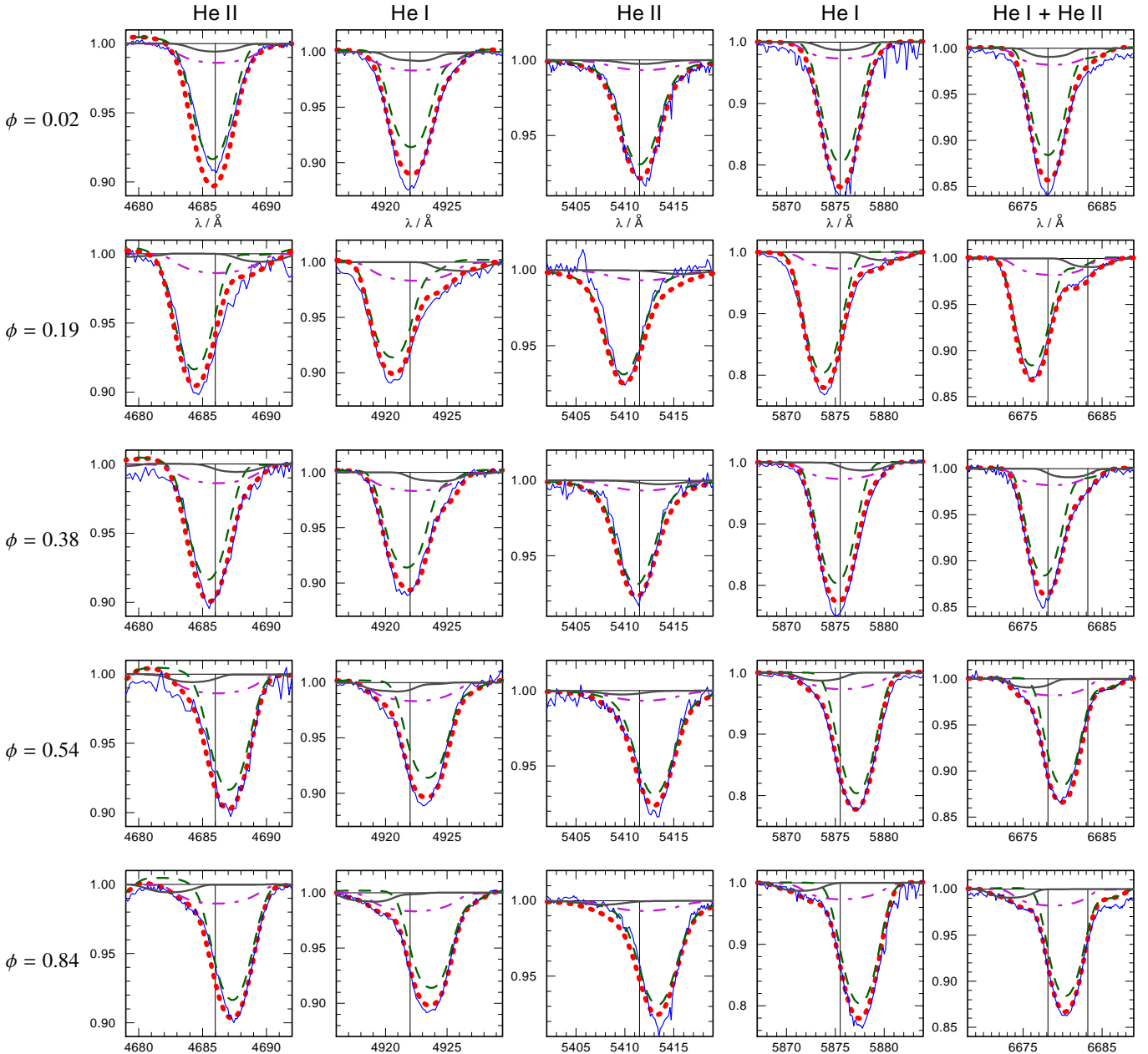}
  \caption{The 25 panels show the contribution of the primary
    (green dashed line), secondary (gray solid line), and tertiary (pink dashed-dotted line) models to the composite
  synthetic spectrum (red dotted line) for six prominent helium lines and five
  different phases, compared with observations (blue line). The lines
  depicted are, from left to right, He\,{\sc ii}\,$\lambda 4686$,  
  He\,{\sc i}\,$\lambda 4922$, He\,{\sc ii} $\lambda 5411$, 
  He\,{\sc i}\,$\lambda 5876$, and the two adjacent lines
  He\,{\sc i}\,$\lambda 6678$ and He\,{\sc ii}\,$\lambda 6683$. The phases are,
  from top to bottom, $\phi = 0.02, 0.19, 0.38, 0.54$, and $0.84$. At each phase, the component models are
  shifted with the velocities given in Table~\ref{tab:shifts}.}
\label{fig:phases}
\end{figure*}

\subsection{Constraining the parameters of Aa2 and Ab}
\label{subsec:contribution}

The five columns of Fig.\,\ref{fig:phases} depict
six prominent He\,{\sc ii}  and He\,{\sc i} lines, 
from left to right: He\,{\sc ii}\,$\lambda 4686$,  
He\,{\sc i}\,$\lambda 4922$, He\,{\sc ii} $\lambda 5411$,
He\,{\sc i}\,$\lambda 5876$, and the two adjacent lines
He\,{\sc i}\,$\lambda 6678$ and He\,{\sc ii}\,$\lambda 6683$. 
Each row depicts a different orbital phase, from top to bottom: $\phi = 0.02, 0.19, 0.38,
0.54, 0.84$. This time, 
we explicitly show the relative contributions
of the primary (green dashed line), secondary (gray solid line) and
tertiary (pink dotted-dashed line) to the 
total synthetic spectrum (red dotted line), compared
to the observations at each phase (blue line). 
In Table \ref{tab:shifts}, we specify 
the RVs with which the three components are shifted at
each phase. The RVs for the primary were inferred in this study and agree very well 
with the RV curve of \cite{Mayer2010} and those of Paper~III.

\subsubsection{The tertiary}

There is a wide and shallow
spectral feature which does not originate in the primary and which 
is constant along all orbital phases (see Figs.\ \ref{fig:phases} and \ref{fig:secondary}).
Like
\cite{Mayer2010}, we identify this feature with the tertiary Ab, which is in fact
the second brightest source in the system. 
The RV of the tertiary, which is practically constant
over all phases, is also inferred independently, and agrees 
well with that suggested by \cite{Mayer2010}. 

We find that the tertiary
contributes significantly more to H\,{\sc i} lines
than to He\,{\sc ii} lines. Together with the visual flux ratio implied from observations, this leads to a
tertiary temperature of $T_* \sim 29.0\,$kK. The gravity and mass-loss 
of the tertiary were constrained based on the Balmer lines. 
As already discussed, it is very hard to identify the
explicit contribution of the tertiary to these lines, and so the
gravity and mass-loss of the tertiary are only roughly
constrained. The parameters derived for the tertiary (cf.\ table \ref{tab:stellarpar}) are consistent 
with it being a $B0~$IV type star \citep{Habets1981, Schmidt-Kaler1982}

\begin{table}
\caption{Radial velocities derived/adopted for each phase in \kms} 
\label{tab:shifts}
\begin{center}
\begin{tabular}{lrrr}
\hline
 Phase & Aa1\footnotemark[1] & Aa2\footnotemark[2] & Ab\footnotemark[1] \\
\hline
0.02 & 0 & 10 & 25 \\
0.19 & -92 & 230 & 25 \\
0.38 & -22 & 120 & 25 \\
0.54 & 83 & -100 & 25 \\
0.84 & 105 & -210 & 25 \\
\hline
\end{tabular}
\tablenotetext{1}{Values for primary and tertiary derived in the analysis}
\tablenotetext{2}{Values for secondary adopted from \cite{Harvin2002}, with the 
  exception of phase $0.84$, which is derived here.}
\end{center}
\end{table}

\subsubsection{The secondary}

\begin{figure}[!htb]
\centering
  \includegraphics[width = \hsize]{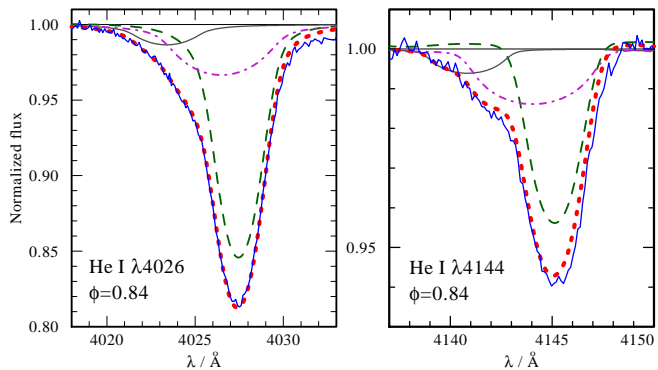}
  \caption{The two panels depict two observed He\,{\sc i} lines (blue lines) at phase $\phi = 0.84$ in which the
    secondary may be detectable. The different curves correspond to the primary (green dashed line),
    secondary (gray solid line), tertiary (pink dashed-dotted line), and total (red dotted line) synthetic spectra.}
\label{fig:secondary}
\end{figure} 

It is very hard, or perhaps impossible, to recognize any contribution from
Aa2 to the spectrum. 
One exception might be the He\,{\sc i} lines $\lambda \lambda 4026, 4144$ at
phase $\phi = 0.84$, which we show in the left and right panels of Fig.~\ref{fig:secondary}, respectively 
(colors and lines are as in Fig.\,\ref{fig:phases}). At this phase,
the primary's RV approaches its maximum of $\sim 110$\,\kms, and so the secondary is expected to be more easily 
observed. The He\,{\sc i} $\lambda4026$ line, for example, seems to have an extended wing towards blueshifted wavelengths. 
The very weak He\,{\sc i} $\lambda4144$ line is one of the few spectral lines which possibly   
portrays a partially isolated feature originating in the secondary. We therefore infer a RV for the secondary
in this phase. For all other phases,
we adopt the secondary's RVs from \cite{Harvin2002}, since the secondary's lines cannot be isolated in the spectrum.
However, as \cite{Mayer2010}
pointed out, it is likely that \cite{Harvin2002} confused the
secondary with the tertiary, so that the secondary RVs reported by
\cite{Harvin2002} are questionable.

Even without directly detecting the secondary, we can still constrain its stellar
parameters. Since the light curve provides a 
lower limit for its visual flux (see 
Sect.~\ref{subsec:assumptions}), the luminosity of Aa2 cannot be 
arbitrarily small. Instead, to avoid too-strong line features from the secondary (which are not observed),  we are forced to change
other stellar parameters, e.g., $T_*$ and $v \sin i$.
The secondary's projected
rotation velocity ($v \sin i \approx 150$\,\kms) agrees with the feature shown in the right panel of
Fig.~\ref{fig:secondary}, and the secondary temperature ($T \approx 26$\,kK)
is consistent with that
obtained from the light
curve analysis of the system (Paper III).

In Sect.~\ref{subsec:assumptions} we argued that the light curve of $\delta$ Ori Aa 
implies that the secondary contributes at least 5.4\%
to total visual flux of the system. In this section, we 
argue that the secondary can contribute no more than this amount.
In other words, our lower
bound for the relative flux contribution becomes also our upper bound, and 
therefore $\Delta V_{\text{Aa1Aa2}} \approx 2\fm8$.
Together with the temperature of the secondary, this enables us to
infer an approximate value for the luminosity of the secondary. Its parameters
suggest it is a $\approx$ B1~V type star \citep{Habets1981, Schmidt-Kaler1982}.

\subsection{Bulk and turbulent motions}
\label{subsec:motions}

\subsubsection{Rotation and macroturbulence}

Rotation is usually the dominant broadening mechanism of photospheric metal lines 
in OB type spectra, as is the case here. 
We infer the projected rotational velocity $v \sin i$  by convolving the synthetic spectrum
with a rotation profile and comparing the line shapes
 to the observed spectrum. While flux convolution is a fair approximation for
photospheric spectra, it
does not
account for the effect of limb darkening and for the extended
formation regions of some lines.
Accounting for rotation in the formal
integration is essential for a consistent inclusion of these effects, but 
is computationally expensive. We therefore use the convolution method
to infer $v \sin i$, and, only after obtaining the best-fit value of $v \sin i$, do we account for 
rotation directly in the formal integration \cite[cf.][]{Hillier2012, Shenar2014}, assuming co-rotation up
of the photosphere ($\tau_{\rm Ross} = 2/3$) and angular momentum conservation in the wind. 
Compared to simple flux convolution, the detailed treatment of rotation generally yields deeper 
lines with less elliptic profiles \cite[e.g.,][]{Unsold}
and may significantly affect the inferred abundances, projected
rotation velocities, and even fundamental stellar parameters.

The effect of rotation on the spectrum
is coupled to the effects of macroturbulence $v_{\rm mac}$, which 
we model using a Radial-Tangential profile \citep{Gray1975, SimonDiaz2007}.
Macroturbulence does not enter the radiative transfer per definition.
Therefore, like solid body rotation, macroturbulence conserves the EWs of the lines. 
Typical values of $v_{\rm mac} \sim 50\,$\kms
are reported for OB type stars \cite[][]{Lefever2007, Markova2008, Bouret2012}. 
While the origin of macroturbulence is not certain, 
it has been suggested to be a manifestation of collective pulsational broadening \cite[e.g.,][]{Aerts2009}. 
$v \sin i$ and $v_{\rm mac}$ are inferred simultaneously from the profile shapes of helium lines and metal lines. 
Not accounting for macroturbulence generally results in 
qualitatively different profiles from what are observed 
in photospheric lines \cite[see e.g., example given by][]{Puls2008}.

\cite{Harvin2002} inferred
$v \sin i = 157\,$\kms for the primary. While this value 
agrees well with the wings of prominent helium and metal lines,
it leads to too-broad Doppler cores. 
A significantly better
fit is obtained with $v_{\rm mac} = 60$\,\kms and 
$v \sin i = 130\,$\kms. For an inclination of 
$i = 76^\circ$ (Paper III) and a radius of $R = 16.5\,R_\odot$ (table
\ref{tab:stellarpar}), this could imply that the rotation period 
is approximately synchronized with the orbital period of $5.7\,$d.

For the tertiary, we find an optimal fit for $v \sin i = 220\,$\kms and $v_{\rm
  mac} = 50\,$\kms, thus confirming the findings of
\cite{Harvin2002} and \cite{Mayer2010} that the tertiary is a rapid
rotator. For the secondary, motivated by 
the arguments discussed in Sect.~\ref{subsec:contribution},
we estimate $v \sin i \sim 150$\,\kms. We adopt $v_{\rm mac} = 0$
for the secondary, lacking any spectral lines from which it can be inferred. Any other 
typical values would bear no effect on our results.

\subsubsection{Microtubulence and terminal velocity}

\begin{figure}[htb]
\centering
  \includegraphics[width = \hsize]{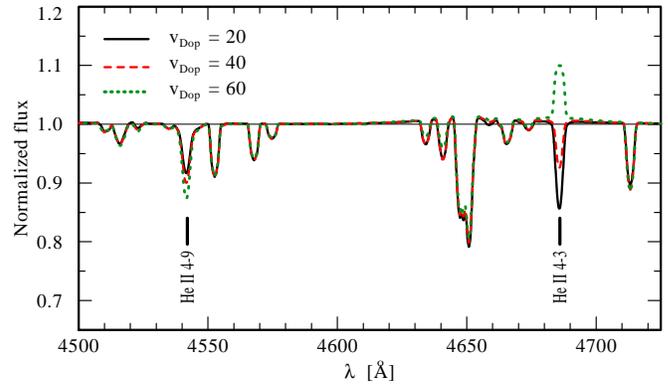}
  \caption{This figure illustrates the extreme sensitivity of the He\,{\sc ii}
  $\lambda 4686$ line to the parameter $v_{\rm Dop}$. Here we depict
  only the primary model, calculated with the parameters in
  Table~\ref{tab:stellarpar}, but with $v_{\rm Dop} = 20$\,\kms (black solid line),
  40\,\kms (red dashed line), and 60\,\kms (green dotted line). The formal integration is
  performed with the same turbulent velocity as given in
  Table~\ref{tab:stellarpar}. Notice that most lines hardly
  react to this parameter.}
\label{fig:vdop}
\end{figure}

In contrast to macroturbulence, microscopic turbulent motion enters directly into the process of 
radiative transfer and generally affects the EWs of spectral lines. 
In the non-LTE iteration, we do not specify the
microturbulence explicitly, but rather the total Doppler width $v_{\rm
  Dop}$ of the opacity and emissivity profiles.
$v_{\rm  Dop}$ thus determines the resolution of the frequency grid in the 
non-LTE iteration.
This parameter generally has a negligible effect on the 
obtained population numbers, with the extreme exception of the He\,{\sc ii} 
$\lambda 4686$ line. As was already noted by \cite{Evans2004}, this line reacts 
very strongly to changes in $v_{\rm Dop}$. We illustrate this in Fig.\,\ref{fig:vdop}, 
where we show a segment of the optical spectrum containing the He\,{\sc ii} 
$\lambda 4686$ line.
In the figure, we depict three models corresponding to the primary calculated with parameters identical 
to those given in Table~\ref{tab:stellarpar}, but with $v_{\rm Dop} = 20$, 
40, and 60\,\kms, respectively. It is evident that the He\,{\sc ii} 4686\,\AA\ line reacts remarkably strongly
to $v_{\rm Dop}$. 
The exact origins of this effect are still under investigation (Shenar et al. in prep.\ 2015), but 
likely involve a feedback effect in the highly non-linear iterative
solution of the radiative transfer problem.
The remaining He\,{\sc ii} lines show a much weaker reaction in the opposite direction. 
Other lines hardly respond to changes of this parameter. 
This example shows that, overall, 
the choice of the parameter $v_{\rm Dop}$ is not critical for the fit, 
and that He\,{\sc ii} $\lambda 4686$
is a poor temperature indicator. Based on this,
$v_{\rm Dop} = 30$\,km/s is used in the analysis, consistent 
with the inferred microturbulence (see below).

In the formal integration, apart from including natural and
pressure broadening, the Doppler width is separated into a thermal
component $v_{\rm th}$, which 
follows the temperature stratification in the model, and a 
depth-dependent microturbulence component $\xi(r)$, which is assumed to
be identical for all ions.
As described in Sect.~\ref{subsec:code}, $\xi(r)$ is interpolated between 
the photospheric $\xi_{\rm ph}$ and wind $\xi_{\rm w}$ turbulent
velocities between two pre-specified radii, all of 
which are free parameters.

\begin{figure}[htb]
\centering
  \includegraphics[width = \hsize]{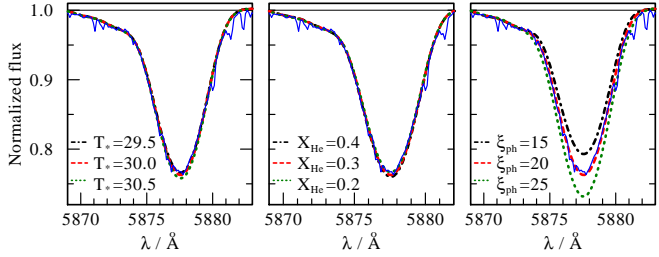}
  \caption{Sensitivity of the He\,{\sc i} $\lambda 5876$ line
    to temperature, helium abundance, and microturbulence. The observed spectrum 
    (blue line) at phase $\phi = 0.84$ is plotted along with three 
    composite synthetic spectra calculated with
    the parameters given in Table~\ref{tab:stellarpar}, but with different 
    temperature (left panel: $T = 29.5, 30,
    30.5$\,kK), helium abundance 
    (middle panel: $X_{\rm He} = 0.2, 0.3, 0.4$), and
    photospheric microturbulence (right panel: $\xi_{\rm ph} = 15, 20,
    25$\,\kms) for the primary model (black dotted-dashed, red dashed, and green dotted
    lines, respectively).}
\label{fig:vmic}
\end{figure}

Values of photospheric microturbulence reported for O giants range between $\sim 10$\,\kms and 
$30$\,\kms \cite[e.g.,][]{Gies1992, Smartt1997, Bouret2012}. 
Since the photospheric microturbulence $\xi_{\rm ph}$ is rarely found to be larger than 30\,\kms, 
its effect on the profile width is negligible compared to the effect of rotation. However, 
$\xi_{\rm ph}$ can have a very strong effect on the EW of spectral lines. The
abundances are thus coupled to $\xi_{\rm ph}$, and wrong turbulence values
can easily lead to a wrong estimation of the abundances
\cite[e.g.,][]{McErlean1998, Villamariz2000}.
To disentangle the abundances and turbulence from, e.g., the temperature, we take
advantage of the fact that different lines respond individually 
to changes in abundances, turbulence, and temperature, depending on
their formation process. Fig.~\ref{fig:vmic} shows an
example. The left, middle, and right panels 
depict the He\,{\sc i} $\lambda 5876$ line as observed at 
phase $\phi = 0.84$ (blue line). In each panel, we show
three different composite synthetic spectra (i.e. composing all 
three components), which were calculated with parameters 
identical to those given in Table~\ref{tab:stellarpar}, except 
for one stellar parameter of the primary. In the left panel, 
$T_*$ is set to 29.5, 30, and 30.5\,kK. In the middle panel, the
helium mass fraction is set to $X_{\rm He}$ of 0.2, 0.3, and 0.4. 
Lastly, in the right-most panel, we set
$\xi_{\rm ph}$ to 15\,\kms, 20\,\kms, and
25\,\kms, respectively.  
The three composite spectra at the left and middle panels can hardly
be distinguished from each other, portraying the insensitivity of 
the He\,{\sc i} $\lambda 5876$ line to temperature and helium
abundance. In the relevant parameter domain, it is 
mainly $\xi_{\rm ph}$
which influences the strength of the He\,{\sc i} $\lambda 5878$ line.
By considering hundreds of lines at all available orbital phases, 
we find that $\xi_{\rm ph} = 20\,$\kms provides the best results
for the primary's model. 
Similarly, we find that a microturbulence of 10\,\kms for the tertiary
yields the best global fit.

\begin{figure}[htb]
\centering
  \includegraphics[width = \hsize]{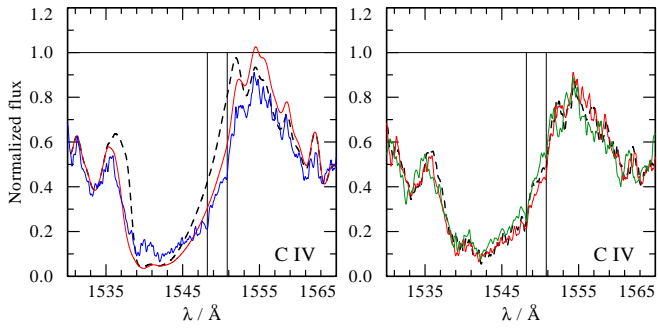}
  \caption{\emph{Left panel:} observed C\,{\sc iv} resonance doublet at phase $\phi = 0.73$ (blue line)
           is compared to two synthetic composite spectra, with increased wind turbulence of $\xi_{\rm w} = 200$\,\kms 
           in the primary model (red solid line), and  
           without it (black dashed line).
           Large microturbulent velocities 
           beyond the sonic point are the only mechanism that can reproduce the redshifted absorption trough. Note that a further improvement of the 
           profile is achieved by accounting for macroclumps, which are not included here.
           \emph{Right panel:} Comparison of three {\it IUE} observations at phases $\phi = 0.18, 0.73, 0.98$ (black, red, and green lines, 
           respectively), illustrating that the redshifted absorption trough is observed at all phases. 
           }
\label{fig:turbwind}
\end{figure} 

We find evidence for a rapid increase of the turbulent velocity in the
primary right beyond the sonic point. The left panel of Fig.\,\ref{fig:turbwind} shows 
the C\,{\sc iv}\,$\lambda \lambda\,1548, 1551$ resonance doublet, as observed in the 
{\it IUE} spectrum taken at phase $\phi = 0.73$. The observation shows a wide absorption trough which extends to red-shifted wavelengths, 
and our task is to reproduce this feature. The black dashed line depicts the 
composite synthetic spectrum calculated with the parameters in Table\,\ref{tab:stellarpar}, but without 
an increased wind turbulence in the primary model, i.e.\ $\xi_{\rm w} = \xi_{\rm ph} = 20\,$\kms. 
The absorption trough is not reproduced. Increasing the terminal velocity only 
affects the blue edge of the line. Varying $\beta$ in the domain $0.7 - 1.5$ does not lead to any notable 
changes in the spectrum. The only mechanism that is found to reproduce the redshifted absorption 
trough is a rapid increase of the microturbulence beyond the sonic point. The red solid line was calculated
like the dashed black line, but with a wind turbulent velocity of $\xi_{\rm w} = 200$\,\kms. 
$\xi (r)$ is assumed to grow from $\xi_{\rm ph}$ at $r_{\rm in} \le 1.1\,R_*$ to $\xi_{\rm w}$  at $r_{\rm out} \ge 2\,R_*$.  At the same time, the blue absorption 
edge is shifted by $\sim 200\,$\kms, thus influencing the value deduced for $v_\infty$. 
The right panel of Fig.\,\ref{fig:turbwind}
shows the same C\,{\sc iv} resonance doublet, 
as observed in three {\it IUE} spectra taken at phases $\phi = 0.18, 0.72, 0.98$ (black, red, and green 
lines, respectively). The figure illustrates the relatively small variability of this line, showing that our results
do not depend on phase, and rejecting the contamination by another component as an explanation for the 
extended red-shifted absorption.
The absorption trough is not reproduced for $r_{\rm out}$ significantly larger than $2\,R_*$, which is
understandable given the need for red-shifted absorption. 
It is interesting to note that it is microturbulence,
and not macroturbulence, which is needed to reproduce this feature.
A further improvement of the line profile fit is obtained by accounting for optically thick clumps (macroclumps) in the wind, 
as we will discuss in Sect.~\ref{sec:macroclump}. 
We do not include macroclumping at this stage in order to single out the effect of 
$\xi_{\rm w}$ on the line profile.

Having inferred the turbulent velocity and
after accounting for clumping in the wind of the primary, 
it is straight-forward to derive the terminal velocity $v_\infty$ from
resonance P Cygni lines. All prominent lines in the UV imply the same
value for $v_\infty$ ($2000 \pm 100$\,\kms).

\subsection{Uncertainties}
\label{subsec:errors}

Since the calculation of a PoWR model
atmosphere is computationally 
expensive, a statistical approach for error determination 
is virtually impossible. 
The errors given in Table\,\ref{tab:stellarpar} for each physical quantity are obtained by 
fixing all parameters but one and varying this parameter, 
estimating upper and lower limits that significantly change the
quality of the fit in many prominent lines relative to the available 
S/N ratio. Errors for parameters which are implied from fundamental parameters via 
analytical relations, e.g., the spectroscopic mass, are calculated by means of error 
propagation. In the case of multiple systems, all models but one 
test model are kept fixed, and only one parameter of the test model is
varied. 

\begin{figure}[htb]
\centering
  \includegraphics[width = \hsize]{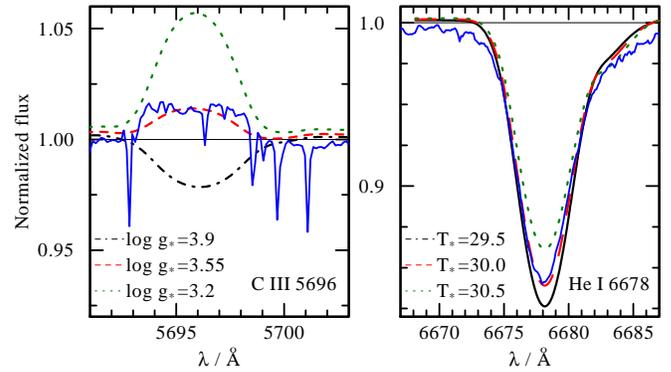}
  \caption{\emph{Left panel:} Sensitivity of C\,{\sc iii} $\lambda5696$ to gravity. The observed spectrum at phase $\phi = 0.02$ (blue line)
           is compared to three composite synthetic spectra calculated with the parameters given in Table\,\ref{tab:stellarpar}, but with $\log g_*$ of the primary set 
           to 3.9, 3.55, and 3.2 (black dotted-dashed, red dashed, and green dotted lines, respectively). 
           \emph{Right panel:} Sensitivity of He\,{\sc i} $\lambda6678$ to temperature. We set the temperature of the primary 
           to 29.5, 30.0, and 30.5\,kK (black dotted-dashed, red dashed, and green dotted lines, respectively).}
\label{fig:errors}
\end{figure}

As mentioned in Sect.\,\ref{sec:results}, it is very hard to constrain the gravity of the primary due to  contamination by the other components.
However, we took advantage of the fact that specific lines do drastically change their strength as a function of gravity. 
An example for such a diagnostic line, C\,{\sc iii} $\lambda5696$, is shown in the left panel of Fig.\,\ref{fig:errors}, as observed at phase $\phi = 0.02$ (blue line).
Three composite spectra (i.e.\ including all components) are plotted, where only the $\log g_*$ of the primary is changed to 3.9 (black dotted-dashed  line), 
3.55 (red dashed line), and 3.2\,\gcgs (green dotted line). 
The remaining stellar parameters are kept fixed to the values given in Table\,\ref{tab:stellarpar}. This line 
only starts to portray emission when $\log g_* \sim 3.5$\,\gcgs. This line also serves as a good example to why $\chi^2$ fitting would not always 
suggest the best fitting model. The contribution of such a small line to the reduced $\chi^2$ is negligible, unlike its diagnostic power.
The right panel of Fig.\,\ref{fig:errors} depicts the sensitivity of the He\,{\sc i} line to the stellar temperature. The temperature of the primary is 
changed to 29.5 (black dotted-dashed line), 30 (red dashed line), and 30.5\,kK (green dotted line). Most He\,{\sc i} and He\,{\sc ii} lines react strongly 
to changes in the temperature and thus enable us to sharply constrain it.

\section{Inhomogeneities in the primary's wind}
\label{sec:macroclump}

Evidence for wind inhomogeneities (clumping) in the winds of hot 
stars are frequently reported.
\cite{Hillier1984} and \cite{Hillier1991_wings} illustrated the effect of
optically thin clumps on the electron scattering wings
of Wolf-Rayet emission lines. More compelling direct
evidence for clumping in the form of stochastic variability  on short timescales
was observed in both Wolf-Rayet \cite[e.g.,][]{Lepine1999}  and OB stars
\cite[][]{Eversberg1998, Markova2005, Prinja2010}. 
Clump sizes likely follow a continuous distribution (e.g., power-law)
which is intimately connected with the turbulence prevailing 
in the wind \cite[e.g.,][]{Moffat1994}. However,
since consistent non-LTE modeling of
inhomogeneous stellar winds in 3D is still beyond reach, the treatment of clumping 
is limited to approximate approaches.

\subsection{Microclumping}

A systematic treatment of optically thin clumps was introduced 
by \cite{Hillier1984} and later by \cite{Hamann1998} using the so-called microclumping approach, 
where the population numbers are calculated in clumps which are a factor of $D$ denser 
than the equivalent smooth wind.  
In this approach, processes sensitive to $\rho$, such as 
resonance and electron scattering, are not sensitive to clumping, and
imply $\dot{M}$ directly.
However, processes which are sensitive to $\rho^2$, such as recombination 
and free-free emission, are affected
by clumping, and in fact only imply the value of $\dot{M} \sqrt{D}$.  
This
enables consistent mass-loss estimations from both types of processes, and 
offers a method to quantify the degree of
inhomogeneity in the wind in the optically thin limit.

\begin{figure}[htb]
\centering
  \includegraphics[width = \hsize]{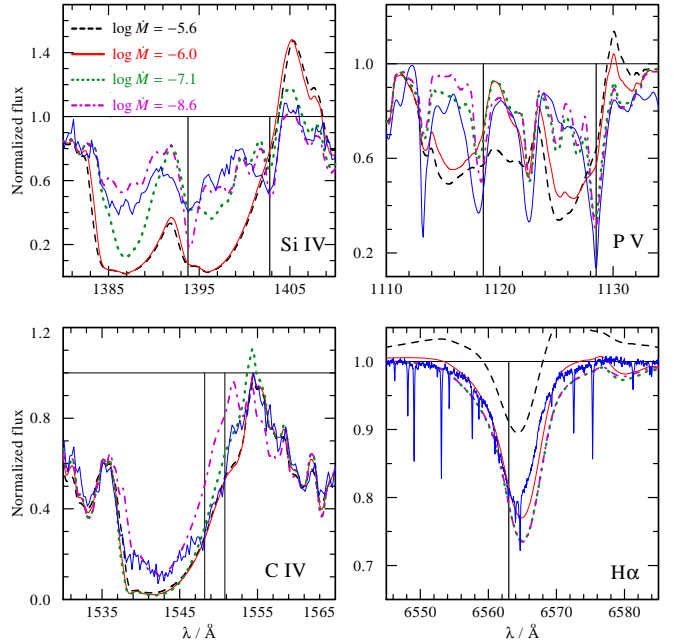}
  \caption{Observed ``wind'' lines (blue line): The Si\,{\sc iv} $\lambda \lambda 1394, 1408$ (upper left), 
    C\,{\sc iv} $\lambda \lambda 1548, 1551$ (lower left), and P\,{\sc v} $\lambda \lambda 1118, 1128$ (upper right) 
    resonance doublets, 
    and H$\alpha$ (lower right), roughly at phase $\phi \sim 0.8$, except for the Copernicus data (P\,{\sc v})
    which are co-added.
    Each panel depicts four composite spectra with the same parameters as in table 
    \ref{tab:stellarpar}, but without clumping ($D = 1$), and with 
    different mass-loss rates: $\log \dot{M} = -5.6, -6.0, -7.1$, and -8.6\,\smy (black dashed, red solid, green dotted 
    , and purple dashed-dotted lines, respectively). The H$\alpha$ and
    different P Cygni lines clearly imply different mass-loss rates, and
    cannot be fitted simultaneously.}
\label{fig:smooth}
\end{figure}

To investigate wind inhomogeneities in the primary Aa1, 
we first adopt a smooth model. 
The four panels of Fig.\,\ref{fig:smooth} depict four 
``wind'' lines: three UV resonance doublets belonging to 
Si\,{\sc iv}, P\,{\sc v}, and C\,{\sc iv} ($\propto \rho$),
and H$\alpha$, as the only recombination line 
potentially showing signs for emission ($\propto \rho^2$). 
In each of the four panels, four composite synthetic spectra (i.e.\ containing
all three components) are plotted
along the observation (blue line). The four models differ
only in the mass-loss rate of Aa1: $\log \dot{M} =$ -5.6 (black dashed line), 
-6.0 (red solid line), \mbox{-7.1} (green dotted line), and -8.6\,$[M_\odot\,{\rm
    yr}^{-1}]$ (purple dotted-dashed line). The remaining stellar 
parameters 
are identical to the ones given in Table \ref{tab:stellarpar} (for $d = 380$\,pc), but 
with $D = 1$ for the primary. The observations roughly correspond 
to phase $\phi \sim 0.8$, except in the case of the P {\sc v} doublet, which 
is obtained from the co-added {\it Copernicus} data (see Sect.\,\ref{sec:obsdata}).

While a best fit for H$\alpha$ (lower right panel in Fig.\,\ref{fig:smooth}) is obtained with
$\log \dot{M} \approx -6.0\,$\,\smy, we
note that  H$\alpha$ portrays noticeable variability 
with time, which increases the uncertainty of the inferred mass-loss rate.
Furthermore, our estimation of the mass-loss rate from
H$\alpha$ could be inaccurate due to underestimated contribution to the
emission by either the tertiary or by wind-wind collision effect.
Fortunately, the H$\alpha$ mass-loss rate is supported by radio
observations. 
Assuming a smooth wind, \cite{Lamers1993} infer a mass-loss rate of $\log \dot{M} \approx
-5.95\,$\,\smy based on the radio flux of
the system for an adopted distance of $d = 500$\,pc.
This value is revised to $\log \dot{M} \approx -6.1\,$\,\smy 
for $d = 380\,$pc,
only slightly smaller than the value derived for H\,$\alpha$ in this study. 
Since both these processes scale as $\rho^2$, and since the wind is assumed to be smooth in 
both cases, we therefore 
conclude that $\log \left(\dot{M} \sqrt{D}\right) \approx -6.0$\,\smy for the primary.

We now turn to the UV resonance lines. 
The Si\,{\sc iv} line, shown in the upper left panel of Fig.\,\ref{fig:smooth},
is clearly not saturated in the observation. In the model, it 
remains saturated for $\log \dot{M} \ge -6.8\,$\,\smy. 
A best fit is obtained with $\log \dot{M} \approx -7.5\,[M_\odot\,{\rm
    yr}^{-1}]$. Together
with the condition discussed in the previous paragraph, 
this implies $D \ge 40$; a best fit is obtained for $D \sim 1000$.
The P\,{\sc v} resonance line (upper right panel)
implies a similar mass-loss rate, and hence a similar clumping contrast.
The C\,{\sc iv} line (bottom left panel), which does not look saturated in the observation\footnote{We 
note that the IUE observation 
may suffer from calibration problems, which
sheds doubt on whether the C\,{\sc iv} resonance line is
unsaturated. However, the observed shape of its absorption trough suggests it is indeed unsaturated. 
This is evident in each of the 60 available {\it IUE} spectra, some even showing this 
more extremely. While we speculate  
that the line is desaturated here, our results do not depend on this strongly.}
, 
requires a much lower mass-loss rate, of the order of $\log \dot{M} \approx
-8.0\,$\,\smy, and as a consequence implies $D \approx 10^4$.

But are such large density contrasts physically sound?
1D and 2D time-dependent hydrodynamic simulations of line driven winds 
suggest typical values of $D = 4- 10$, 
with $D = 20 - 100$ occurring in the most extreme cases
\cite[][]{Owocki1988, Feldmeier1997, 
Runacres2002, Sundqvist2013}. 
Evolutionary considerations support $D \sim 2-3$ in order to obtain sufficient 
mass-loss from OB type stars \cite[e.g.,][]{Hirschi2008}. 
Values of $4 - 10$ are typically reported for Wolf-Rayet stars 
based on electron scattering wings of strong emission lines 
\cite[e.g.,][]{Todt2013, 
Hainich2014},
while larger density contrasts of the order of $20$ or more are suggested for OB stars \cite[e.g.,][]{Bouret2012}
to reconcile the strong discrepancy between $\dot{M}$ values derived from \mbox{P Cygni} resonance lines 
and recombination lines. Analyses of the P\,{\sc v} resonance doublet
simultaneously with optical recombination lines \cite[e.g.,][]{Fullerton2006} may imply clumping factors 
as large as $100$. However, we note that the latter authors neglect the effect of porosity
\cite[][]{Oskinova2007}, and that
the resulting low mass-loss rates are inconsistent with 
polarization studies based on electron scattering \citep{St-Louis2008}, 
which depends linearly on density. It seems therefore 
that the implied values for $D$ exceed plausible limits, 
and that no single value for $D$ can satisfy all emission features simultaneously.

One possibility is that the abundances, or any other 
fundamental parameters derived for Aa1 which may affect the formation of the UV resonance lines, 
are significantly inaccurate. This is unlikely, because 
the photospheric features, as well as the general shape of the UV iron forest, are 
well reproduced and do not leave room for variations. Another possibility
is that one of the components, perhaps through interaction with the primary's wind, is contaminating
the observation. However, there are no clear indications for a periodic variation of 
the UV resonance lines with phase, as could be expected in such a case.
We therefore suggest that not only is the wind of Aa1 clumped, but also 
that the clumps are optically thick in the strong UV lines.
As illustrated by \cite{Oskinova2007},
the optically thickness of 
the clumps leads to an effective reduction of the opacity of strong UV resonance lines, 
and could thus enable
to obtain results which are consistent over the X-ray, UV, optical, IR and radio regimes.
A further indication 
for optically thick clumps comes from the variable EW ratio of the of the two
Si\,{\sc iv}
components. 
A ratio of 1 is expected to occur 
in the extreme case where the porous wind consists of optically thick clumps and 
``holes'' between them, while a ratio of $\approx2$
(corresponding to the ratio of oscillator strengths of the two components)
should occur in the case of a homogenous wind. 
The EW ratio of the Si\,{\sc iv} doublet,
which is observed to be smaller than 2, could be explained 
by the presence of optically thick structures present in the wind 
\cite[cf.][]{Prinja2010}.

\subsection{Macroclumping}

In PoWR, macroclumping is implemented only in the formal integration,
leading to an effective reduction of the opacity in strong lines.
\cite{Oskinova2007} thoroughly discuss the method,
and illustrate the significant effect of
macroclumping in the O supergiant $\zeta$ Puppis.
\cite{Surlan2013} model the effect of macroclumps by
means of 3D Monte Carlo simulations and obtain similar results.
A consequence of accounting for macroclumps is the need to introduce a  
further parameter, 
$L_{\rm mac}$, which specifies the separation between the clumps.
The non-LTE nature of the line formation makes a simultaneous prediction of 
$L_{\rm mac}, D$, and $\dot{M}$  
practically impossible. We are therefore forced to adopt a value for one of these
parameters. Since not much is known about the geometry of the clumps, we avoid
prespecifying $L_{\rm mac}$. Instead, we choose to adopt $D = 10$ as 
a compromise between $D \sim 2$ and $D \sim 100$. Motivated by hydrodynamic studies \cite[][]{Feldmeier1997, 
Runacres2002}, we assume that the clumping initiates
at $r = 1.1\,R_*$ and grows to its maximum contrast of $D=10$ at $r \sim 10\,R_*$. 
Our results depend  
only weakly on the depth-dependence of $D(r)$.

\begin{figure}[htb]
\centering
  \includegraphics[width = \hsize]{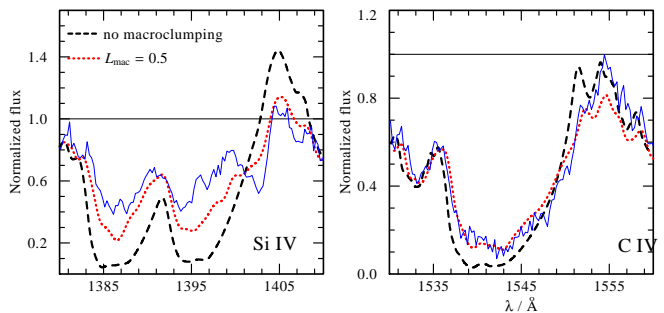}
  \caption{The observed Si\,{\sc iv} and C\,{\sc iv} resonance doublets 
(left and right panels, respectively) 
          at phase $\phi = 0.83$ (blue line) compared to the composite 
synthetic spectrum calculated with the parameters 
          in Table\,\ref{tab:stellarpar}, but without macroclumping in the 
primary (black dashed line), and with macroclumping using $L_{\rm mac} = 0.5\,R_*$ 
(dotted red line).}
\label{fig:macro}
\end{figure}

Fig.\,\ref{fig:macro} portrays the two C\,{\sc iv} and Si\,{\sc iv} resonance doublets as 
observed at phase $\phi = 0.83$ (blue line). 
The black dashed line depicts the synthetic composite spectrum without the inclusion 
of macroclumping in the primary. The red 
dotted line depicts the same composite spectrum, but with $L_{\rm mac} = 0.5\,R_*$ adopted for the primary. 
The strong effect of macroclumping
on the resonance lines is evident. The $H\alpha$ line, as well as the photospheric features, are hardly affected 
by the macroclumping formalism. 
The value $L_{\rm mac} = 0.5\,R_*$ provides a fair compromise for most wind lines, but the Si\,{\sc iv} 
resonance lines 
in fact require larger values of the order of $L_{\rm mac} \sim \,R_*$. The analysis therefore suggests 
$L \gtrsim 0.5\,R_*$. 
This value should not be given too
much significance, as macroclumps are treated only as a 
rough approximation here. 
Nevertheless, we accounted for the major effects expected to rise 
from both optically thick and optically thin clumps, so mass-loss rates
are unlikely to be very different than those derived here.
Future variability studies should help to further constrain 
the amount of inhomogeniety in the primary's wind.

After assuming a clumping factor, the mass-loss rate is fairly well constrained, but
still depends on the adopted distance. For the adopted distance of $d = 380$\,pc, 
the mass-loss rate is found to be $\log
\dot{M} = -6.4\pm0.15$\,\smy. 
Since the mass-loss rate scales as $D^{-1/2}$, we can consider the two extreme alternatives 
for the clumping factor $3 \le D \le 50 $ to set lower and upper bounds for the mass-loss rate
of the primary: \mbox{$-6.2 \le \log \dot{M}_1 \le -6.8$\,\smy}. 
Interestingly,
\cite{Vink2000} predict $\dot{M}_{\text{Vink}} = -6.48$\,\smy for a star with the 
parameters of the primary (for $d = 380$\,pc) based on hydrodynamic 
calculations, which is in very good agreement 
with our results for the adopted clumping factor of $D = 10$ (cf.\ Table \ref{tab:stellarpar}). 
However, we note that \cite{Vink2000} performed their calculations 
for homogenous winds, and it is not clear whether their predictions would
remain the same 
in the case of significant inhomogeneities \citep{Muijres2011}.

\section{Where do the X-rays in $\delta$~Ori~A come from?}
\label{sec:xrays}
So far, our analysis focused on the ``cool''  stellar wind. However, 
the cool wind alone cannot account for the 
observed X-rays in $\delta$~Ori~A. 
It is commonly believed that X-ray emission in single stars originates in the wind
due to instability of the line-driving mechanism \cite[e.g.,][]{Feldmeier1997} 
or via acoustic driving from subsurface convection \cite[][]{Cantiello2009}.
In binary systems, an excess of hard X-ray flux may originate from wind-wind collisions
\cite[][]{Williams1990, Corcoran2003}. 
In this section, we explore the properties of the ``hot'' X-ray 
producing component.

\subsection{Auger ionization}
\label{subsec:auger}

The presence of strong X-ray radiation in OB type stars was
hypothesized prior to the first direct X-ray observations in massive stars.
\cite{Cassinelli1979} were the first to suggest that the detection of
UV resonance lines of high-ionization ions such as N\,{\sc v} and 
O\,{\sc vi}, which are observed in many O type stars, 
may indicate that Auger ionization \cite[][]{Meitner1922, Auger1923}
plays an important role in their formation. Auger ionization
occurs when very energetic photons remove an electron of an inner shell 
(usually the K-shell), ultimately resulting in a double ionization.
The Auger effect on UV spectra has
been frequently 
detected in studies of OB type stars \cite[e.g.,][]{Oskinova2011}.

The effect of 
X-rays in the wind of the primary Aa1 is evident by merely inspecting its UV spectrum. 
The presence of the strong UV resonance doublets
O\,{\sc vi} $\lambda\lambda1032,1038$ and N\,{\sc v} $\lambda\lambda1239,1243$
cannot be reproduced otherwise. 
Indeed, an X-ray luminosity of $\approx1.4\cdot10^{32}$\,erg\,${\rm s}^{-1}$ is reported 
for $\delta$ Ori A  (Paper I), which corresponds to $\sim
10^{-6.87}$ times the total bolometric luminosity of the system\footnote{We use here the 
total bolometric luminosity because all components are OB type stars and are thus expected 
to emit X-rays proportionally to their luminosities.}.
The question remains, however, as to the origin of the X-ray 
radiation observed.

There are several arguments which suggest that emission from
wind-wind collisions do not dominate the X-ray flux in $\delta$ Ori A. 
Paper~II reports a relatively
weak variability of the X-ray flux, which they cannot tie 
with certainty to phased-locked variations, in particular wind-wind collisions. 
The inferred value of $\log L_{\rm X} / L_{\rm Bol} \sim -6.87$ is 
typical for OB type stars \cite[e.g.,][]{Pallavicini1981, Seward1982, Moffat2002, Oskinova2005,
Naze2009}, and does not imply any excess X-ray radiation from strong wind-wind collision.
Lastly, the N\,{\sc v} $\lambda\lambda1239,1243$ resonance doublet, which clearly forms
due to the X-ray radiation, is persistent throughout all available {\it IUE} spectra 
(see right panel of Fig.\,\ref{fig:auger} below).
If most X-rays originated in a 
collision zone which irradiated only 
a part of the star (whose orientation relative to the observer depended on phase), 
then a larger variability could be anticipated. 
It therefore seems plausible to assume that the X-rays 
originate in the wind itself.

We model the effect of X-ray radiation as described by \cite{Baum1992}. The X-ray
emission is assumed to originate in optically thin filaments of shocked plasma
embedded in the wind. 
The X-ray field is characterized 
by the temperature $T_{\rm X}$ of the shocked plasma and the radially-constant filling
factor $X_{\rm fill}$, which describes the ratio
of shocked to non-shocked plasma.
The onset radius of shocked plasma is denoted by $R_0$. The
X-ray emissivity $\eta_{\rm X}(r)$ at each radial 
layer $r > R_0$ is proportional to $\rho^2$.
In principle, the three parameters $T_{\rm X}$, $X_{\rm fill}$ and $R_0$ are chosen such that the observed X-ray SED
(Huenemoerder, priv. com.) is approximately reproduced by the synthetic X-ray SED emerging from the model,
after accounting for interstellar extinction. Once 
the onset radius $R_0$ has been fixed, $T_{\rm X}$ and $X_{\rm fill}$ follow from the observed SED. 
Motivated by the f/i analysis and X-ray line modeling results (see Sects.~\ref{subsec:xraysmod} and \ref{subsec:fir}), we fix the
onset radius to $R_0 = 1.1\,R_*$, leading to $T_{\rm X} = 3\,$MK and $X_{\rm fill} = 0.1$. These are only rough 
approximations to the X-ray properties in the wind (see Paper~I), used to reproduce the bulk of X-ray emission observed.

\begin{figure}[!htb]
\centering
  \includegraphics[width = \hsize]{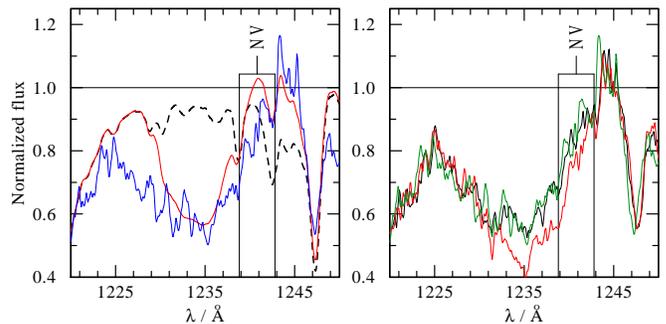}
  \caption{\textit{Left panel:}  Observed N\,{\sc v} resonance doublet 
           at phase $\phi \sim 0.1$ (blue line) with two synthetic composite
           spectra, calculated with the parameters in Table\,\ref{tab:stellarpar}, 
           not including X-rays (black dashed line), and including X-rays (red solid line).
           \textit{Right panel:} Like the right panel of Fig.\,\ref{fig:turbwind}, the three 
           {\it IUE} observations at phases $\phi = 0.8, 0.3, 0.1$ (black, red, and green lines, respectively).}
\label{fig:auger}
\end{figure}

The left panel of Fig.\,\ref{fig:auger} shows the effect of including X-rays in the wind of the primary on the N\,{\sc v} resonance doublet
$\lambda\lambda1239, 1243$. The blue line depicts the {\it IUE} observation of this doublet at phase $\phi = 0.1$. The black dashed
line plots the synthetic composite spectrum without the inclusion of X-rays.  
Since the temperature in the wind is far from being 
sufficient to populate the N\,{\sc v} ground state, only photospheric absorption is obtained. The red solid line is obtained after the inclusion of X-rays. 
While the line shape is not accurately reproduced, it is clear that 
X-rays are required in order to reproduce the observed P Cygni N\,{\sc v} resonance line. Note also 
that the line is blended with components of iron lines in the blue part, making a determination 
of its terminal width difficult.
The right panel of Fig.\,\ref{fig:auger} depicts three {\it IUE} 
observations at phases $\phi \sim 0.8, 0.3, 0.1$. While the N\,{\sc v} resonance doublet shows stronger 
variability than most other resonance lines, the P Cygni feature is persistent and clearly visible in 
all 60 available {\it IUE} spectra.

\subsection{f/i analysis}
\label{subsec:fir}

Spectra of He-like ions (i.e. ions with two electrons) 
exhibit a group of three neighboring X-ray lines referred to as resonance (r),
forbidden (f), and intercombination (i) lines. 
A well established method to constrain the formation region of X-rays in stellar 
winds makes use of observed 
forbidden-to-intercombination (f/i) line ratios.
f/i analyses were originally developed to 
study the solar X-ray radiation \cite[][]{Gabriel1969, 
Blumenthal1972}, but 
are now also frequently used to study the X-ray emission of 
OB and Wolf-Rayet stars \cite[e.g.,][]{Waldron2001, Leutenegger2006, Waldron2007}. 
The f/i line ratio is determined by the relative population of the upper levels of the f and i 
lines, altered either by collisions 
or by photo-excitations \citep[e.g.,][]{Porquet2001}. 
For each helium-like ion, there are three possible transitions between the upper levels of the f and i lines, denoted in the following with indices $j = 0,1,2$. 
These transitions
are typically at UV frequencies, although their exact wavelengths depend on the ion.
The stronger the UV radiation field is, the smaller the f/i ratio becomes. 
A detailed treatment allows one to construct an equation which
predicts the f/i line ratio $\mathcal{R}$ as a function of the radiative
excitation rate $\phi$ and electron density $n_{\rm e}$ \cite[cf.][Eq.\,1c]{Blumenthal1972}.

The interpretation of observed f/i line ratios requires some discussion. After all, what we observe 
is likely the X-ray radiation reaching us from an extended region where it is formed.
The simplest way to interpret the observations would be to assume that
the X-ray emitting gas is distributed over a thin spherical shell  
located at a formation radius $R_{\rm form}$, often referred to as the point-like interpretation. 
However, we know from detailed X-ray line fitting (see Sect~\ref{subsec:xraysmod}) 
that the X-ray radiation must originate in an extended region. Obviously, the point-like interpretation cannot describe the whole truth. 
A more generalized interpretation to the observed f/i 
line ratios, thoroughly described by \cite{Leutenegger2006}, involves the integration of the X-ray 
radiation emanating from a continuous range of radii. Within the
assumptions of this method, it is not the formation region which is sought,
but the \emph{onset} radius $R_0$ of the X-ray emission. 
We refer the reader to studies by 
\cite{Gabriel1969}, \cite{Blumenthal1972}, and \cite{Leutenegger2006} for
a detailed description of the methodology of both interpretations.

We do not calculate the radiative excitation rate of the
upper $f$ level
by diluting the photospheric fluxes, \cite[cf.][Eq.\,2]{Leutenegger2006}, 
but instead use directly the mean intensities $J_\nu$ at each radial layer, as obtained by our PoWR model. 
This way, we account for diffuse emission and limb darkening in a
consistent manner. We include all  three transitions $j = 0, 1,
2$ in the calculation, properly weighted with their respective branching ratios.
The relevant wavelengths $\lambda_j$ and oscillator strengths $f_j$ are
extracted from the NIST database.

\begin{figure}[htb]
\centering
  \includegraphics[width = \hsize]{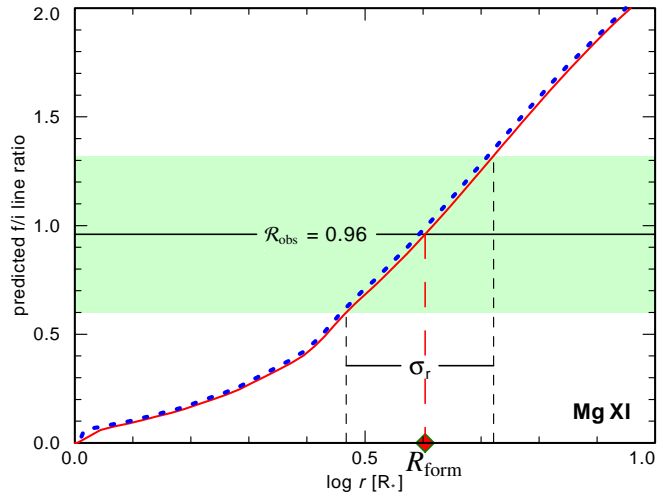}
  \caption{Theoretical f/i ratio $\mathcal{R}(r)$ is plotted as a function of the X-ray formation radius (blue dotted line)
           for the Mg\,{\sc xi} ion. The intersection of $\mathcal{R}(r)$ with the observed value 
           $\mathcal{R}_{\rm obs}$ implies the formation radius under the assumptions of the point-like
           approach. The error $\sigma_{\rm r}$ in the formation radius corresponds to 1-$\sigma$ measurement 
           uncertainties in $\mathcal{R}_{\rm obs}$, depicted by the green shaded rectangle. 
           The red solid line depicts $\mathcal{R}(r)$ as well, but includes the contribution of 
           collisions assuming a full ionization and 
           a factor 1000 density enhancement in the shocked plasma
           }
\label{fig:eldens}
\end{figure}

Fig.\,\ref{fig:eldens} shows an example calculated for the He-like Mg\,{\sc
  xi} ion. The blue dotted line depicts the f/i ratio
$\mathcal{R}(r)$ predicted by the model for the point-like assumption 
as a function of the formation radius $R_{\rm form}$, neglecting
collisional excitation. The solid
horizontal line depicts
the observed value of $\mathcal{R}_{\rm obs} = 0.96\pm0.36$ (Paper I). The shaded green area depicts the 
1-$\sigma$ measurement uncertainty. 
If the major part of the X-ray radiation originates at one radial layer, 
this layer will be
located at $R_{\rm form} = 4.1^{+1.2}_{-1.0}\,R_*$, where the uncertainty
$\sigma_{\rm r}$ corresponds to the measurement uncertainty.
In Fig.\,\ref{fig:eldens}, we also illustrate the influence of
collisional excitation on $\mathcal{R}(r)$. The red solid line also plots 
$\mathcal{R}(r)$, but accounts for collisional excitation, assuming a full ionization and 
an unrealistically large factor of 1000 for the density enhancement in the shocked regions. Evidently, the contribution of collisions to the excitation of
the upper $f$ level is negligible in $\delta$~Ori~A.

\begin{figure}[htb]
\centering
  \includegraphics[width = \hsize]{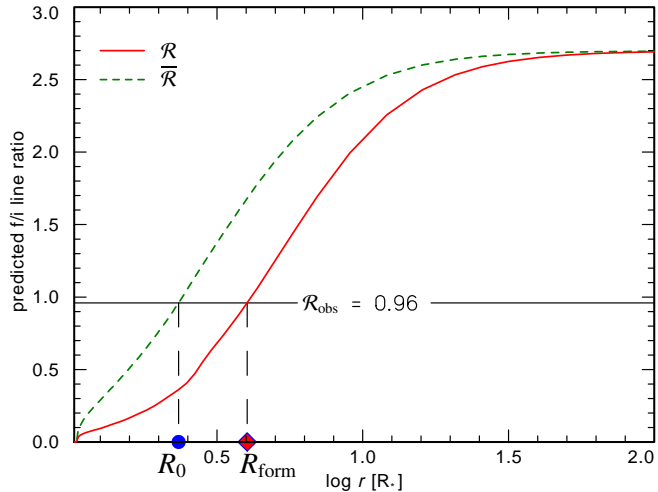}
  \caption{Same as Fig.\,\ref{fig:eldens}, but now accounting for an extended region 
           of X-ray formation (green dashed line). The intersection of $\mathcal{\overline{R}}(r)$ with the 
           observed value delivers the onset radius $R_0$.}
\label{fig:integrated}
\end{figure}

In Fig.\,\ref{fig:integrated}, we compare the point-like interpretation with
the more generalized interpretation described by \cite{Leutenegger2006} for
the ion Mg\,{\sc xi}. The red solid line depicts $\mathcal{R}(r)$, as in
Fig.\,\ref{fig:eldens}. The green dashed line shows the predicted f/i ratio 
$\mathcal{\overline{R}}(r)$ as a function of the \emph{onset} radius
$R_0$. Note that while both functions are depicted in the same coordinate
frame, the meaning of $r$ is different. As is evident from the figure,
at the observed value of $\mathcal{R} = 0.96$, the onset radius $R_0$ predicted by
integrating and weighting the contribution of all layers at $r > R_0$ is
approximately $2.4\,R_*$.

\begin{figure}[htb]
\centering
  \includegraphics[width = \hsize]{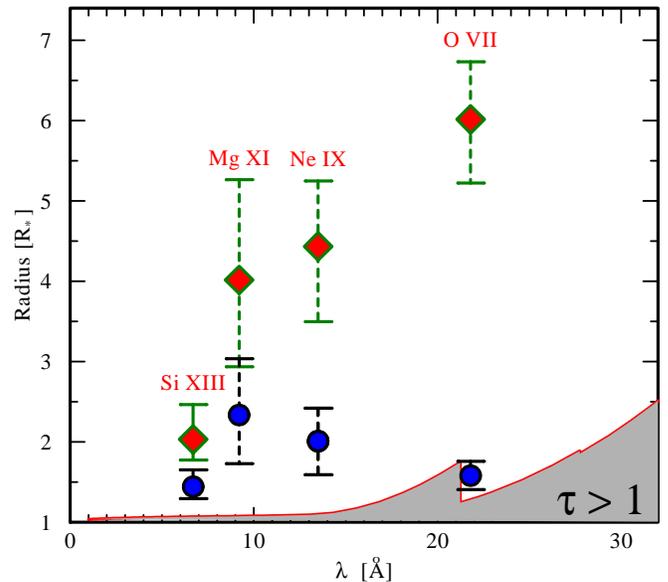}
  \caption{Inferred formation radii (red diamonds) and 
          onset radii (blue circles) of the X-ray radiation, as inferred from the four 
          He-like ions Si\,{\sc xiii}, Mg\,{\sc xi}, Ne\,{\sc ix}, and O\,{\sc vii}. 
          The error 
          bars correspond to measurement uncertainties propagated into errors of 
          the inferred radii (see Fig.\,\ref{fig:integrated}). The shaded area corresponds 
          to radii at which $\tau_{\rm Ross} \ge 1$.
           }
\label{fig:firresults}
\end{figure}

Finally, Fig.\,\ref{fig:firresults} graphically
summarizes our results 
for the He-like ions Si\,{\sc xiii}, Mg\,{\sc xi}, Ne\,{\sc ix}, and O\,{\sc
  vii} using both the point-like and generalized interpretations. The measured
f/i ratio of S\,{\sc xv} does not provide any constraints, so this ion is not included.
Measured f/i values and their uncertainties are given in Paper I. 
Formation radii inferred for each line are indicated with red diamonds, 
while onset radii are indicated with blue circles. The dashed vertical lines 
mark the $1 - \sigma$ error range corresponding to f/i measurement uncertainties. The shaded 
gray area in both panels depicts the optically thick region in the model, i.e.\ 
regions where photons cannot escape.

Evidently, the X-ray formation radii are different for ions with 
significantly different ionization potentials (e.g.,\ Si\,{\sc xiii} vs. O\,{\sc vii}). 
This may suggest that a distribution of temperatures governs the formation radii 
\citep[see e.g.,][]{Krticka2009, Herve2013}. Fig.\,\ref{fig:firresults} suggests that 
higher ions are formed at lower radii. This could imply that the 
hotter plasma is found closer to the stellar surface, while cooler plasma dominates 
farther out, a fact which was also pointed out by \citet{Waldron2007} from their 
study of a large sample of O-type X-ray spectra. The correlation between the ionization 
potential and location in the wind  is no longer seen for the onset radii, which  
may suggest that X-rays are emitted in all He-like lines already 
very close to the stellar surface, independent of the element. This is in agreement 
with a picture based on hydrodynamic simulations, where plasma of different temperatures 
is present in cooling layers behind a shock front \citep{Feldmeier1997}. It is also 
worth noting that the formation radii roughly follow the same trend as the $\tau=1$ surface
of the cool wind, providing an independent confirmation that our analysis 
provides realistic values for the cool wind opacity.

\subsection{X-ray line modeling}
\label{subsec:xraysmod}

We now demonstrate that the high-resolution X-ray spectrum  of $\delta$
Ori~A can be described consistently with the cool stellar  wind model
obtained from the analysis of UV and optical spectra. We use the high 
signal-to-noise ratio X-ray spectra  obtained by the {\em Chandra}'s 
HEG and MEG detectors.  The
observation consists of four different exposures taken at  different
orbital phases. 
A complete description of the X-ray observations and data
analysis is given in Paper I. 
During the orbital motion, the radial  velocity of the
primary changes between $+114$\,km\,s$^{-1}$ and $-78$\,km\,s$^{-1}$, 
with a systemic velocity of $\gamma = 15$\,\kms
(Paper III). It is ambiguous whether the observed X-ray 
Doppler shifts follow the expected orbital pattern (Paper II). 
Regardless, our tests show that the resulting 
Doppler shifts, which are comparable to the instrumental resolution 
and are much smaller than the wind velocity, bear a negligible effect 
on the modeled line profiles. We therefore neglect Doppler shifts 
due to orbital motion, and model the line profiles at the systemic 
velocity of $15$\,\kms.  For all L$\alpha$ lines
of H-like ions, both components of the  doublet are taken into account
in the model.  

We analyze the X-ray spectra by simulating X-ray lines using our 2-D 
stochastic stellar wind code \citep{Oskinova2004} and comparing them 
with observed lines. As we are interested in the line profiles, 
the maximum of the synthetic lines is normalized to their observed
maximum.  Our model is based on the assumption of a two  component
medium: the ``cool wind'' and the ``hot wind''. The X-ray  radiation 
is assumed to originate in optically thin filaments 
evenly distributed in the hot wind component 
\citep[the treatment of resonant scattering in X-ray lines is presented in][]{Ignace2002}.  
The emissivity $\eta_{\rm X}$ scales as $\rho^2$, and the filling factor 
is assumed to be radially constant. A more sophisticated assumption on the behavior 
of $\eta_{\rm X}$
could lead to slightly different results numerically, but should not affect 
our conclusions qualitatively. The X-rays are attenuated on their way 
to the observer due to K-shell  absorption in the cool wind. Hence the X-ray 
propagation  in the stellar wind is described by a pure absorption case of 
radiative  transfer and is therefore relatively simple 
\citep[e.g.,][]{Ignace2001, Oskinova2004}.

\citet{Macfarlane1991} demonstrated that the shape of optically thin 
emission lines is sensitive to the column density of the cool absorbing 
material: the lines become more skewed to the blue when the cool wind column 
density is large. They suggested using this property to measure the cool wind 
density. However, in clumped winds, the wind column density is 
reduced \citep{Feldmeier2003}, implying that the analysis of observed X-ray emission line 
profiles in O-type stars should account for wind clumping.

Several previous studies of X-ray emission lines in the spectra of 
O-type stars reported that accounting for wind clumping in the 
models does not improve the quality of the fit to the X-ray spectrum 
\citep[e.g.,][]{Herve2013}.  This is likely due to a degeneracy between 
the effects of mass-loss and clumping on the X-ray spectrum: 
one can always neglect clumping, albeit at the cost of lower 
estimated wind opacity. However, as we show here, the effect 
of clumping cannot be assumed to be negligible for the realistic 
mass-loss rate derived from the detailed analysis of the cool wind.

In this study, we extract the mass-absorption coefficient $\kappa_\lambda(r)$        
directly from the PoWR model of the primary's cool wind.
Importantly, our models include clumping and the feedback of 
X-ray radiation on the ionization structure of the wind and  
are therefore consistent. We make a simplifying assumption that the velocity 
field of the hot X-ray emitting plasma is the same as that of the cool X-ray 
absorbing wind. Thus, with the exception of the radial distribution of the hot 
plasma, all X-ray model parameters are determined by the non-LTE stellar atmosphere 
model.    

The theory of X-ray transfer in clumped stellar winds was developed 
by \citet{Feldmeier2003} and \citet{Oskinova2004}. The special case of  
spherical clumps was considered by \citet{owocki2006}. The macroclumping formalism 
does not make any {\it ad hoc} assumptions about the size and optical depths of the 
cool wind clumps and is therefore suitable for effectively describing both smooth 
winds as well as clumped winds. This formalism further allows one to treat 
angle-dependent opacities, e.g.,\ non-spherical wind clumps. 
A key parameter of the macroclumping formalism is the {\it fragmentation frequency}
$n_0$ [s$^{-1}$] -- the number of clumps per unit time passing 
through a sphere at an arbitrary radius, which does not depend on the radius due 
to the assumption of clump conservation. $n_0$  is approximately related to the 
average separation between clumps $L_{\rm mac}$ (see Sect.\,\ref{sec:macroclump}) 
via $L_{\rm mac}^3\approx R_\ast^2 v_\infty n_0^{-1}$. Allowing for an 
angular-dependent opacity, e.g.,\ flattened clumps or shell fragments, results 
in distinct line shapes, with  flat topped line profiles in the limiting 
case of opaque clumps \citep[see also][]{Ignace2012}. For each line, we consider a 
smooth model and a clumped model with anisotropic wind opacity \cite[cf.][]{Oskinova2004}, 
assuming flattened, ``pancake-like'' clumps \citep{Lepine1999}.
Comparing the observations with a grid of inhomogenious models, 
we find that $n_0 = 8.6\times 10^{-5}$\,s$^{-^1}$ provides a good compromise for several 
X-ray lines. This roughly corresponds to a wind flow time of 
$t_{\rm flow} = R_* / v_\infty \approx 1.5$\,h ($n_0 \approx t^{-1}_{\rm flow}$),
and to $L_{\rm mac} \sim 1\,R_*$, consistent with the lower bound $L_{\rm mac} \gtrsim 0.5$
implied from the UV and optical analysis. 

\begin{figure}[htb]
  \includegraphics[width=\columnwidth]{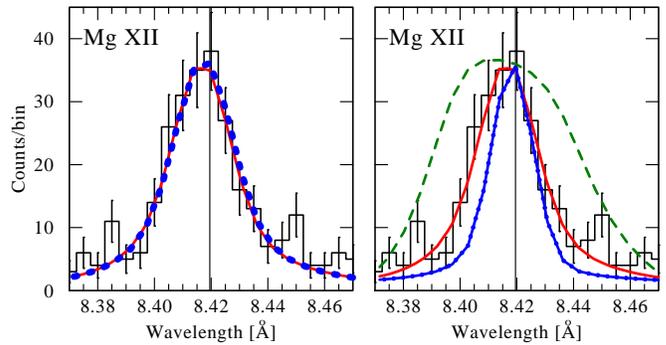}
\caption{
{\it Left panel}: Observed co-added HEG$\pm 1$ spectrum of \dori\ centered on the
Mg\,{\sc xii} line (histogram), compared with a smooth-wind model (blue dotted line), 
and a clumped model (red solid line). As anticipated, 
clumping hardly affects the line formation at short wavelengths.
{\it Right panel}: Again
Mg\,{\sc xii}, with three smooth models which assume X-ray 
onset radii of $1.01\,R_*$ (blue dotted-solid line), $1.1\,R_*$ (red solid line) and $1.5\,R_*$ (green dashed line), suggesting 
that X-ray radiation \emph{initiates} roughly at $1.1\,R_*$ (for the adopted velocity law). The vertical lines indicate the rest 
wavelength shifted with the systemic velocity $15$\,\kms
}
\label{fig:onsetrad}
\end{figure}

The cool wind is virtually transparent at wavelengths shorter 
than $10$\,\AA\ (see Fig.\,\ref{fig:firresults} in Sect.\,\ref{subsec:fir}). 
In this case, the line shape can provide vital information regarding 
the velocity field and density distribution of the hot plasma.
The left panel of
Fig.\,\ref{fig:onsetrad} shows the observed  Mg\,{\sc xii}
line compared with a
smooth-wind  model (blue dotted line) and a clumped-wind model (red solid
line).  
The Mg\,{\sc xii} line 
has the highest S/N ratio in the HEG wavelength range, and since 
the HEG has a superior spectral resolution (0.012\,\AA)
compared to the MEG (0.023\,\AA), it is especially useful for
studying the detailed shape of line profiles.  
Fig.\,\ref{fig:onsetrad} illustrates the above statement: lines at $\lambda
\lesssim 10$\,\AA\ are hardly absorbed in the wind, and thus do not show any
sensitivity to wind clumping. Furthermore, the good agreement between the 
models and the observations
implies that the majority of X-rays originate in filaments co-moving with 
the cold wind, exhibiting the whole range of velocities up to the 
roughly the terminal velocity in their line profiles.

The right panel of
Fig.\,\ref{fig:onsetrad} shows a comparison  between the observed
Mg\,{\sc xii} line and three smooth-wind models that assume different 
onset radii for the X-ray emission: 1.01\,$R_*$ (blue dotted solid line), 1.1\,$R_*$ 
(red solid line), and
1.5\,$R_*$ (green dashed line).  This comparison clearly illustrates that
X-rays must form already very close to the stellar surface, at around $1.1\,R_*$. Larger onset
radii result  in flat-topped line profiles which do not agree at all
with the observed profiles. Smaller onset radii imply narrower lines than those observed.  
We note that the exact values are very sensitive to the adopted velocity law. However, the same conclusion is obtained 
for all our test models, where we vary the exponent of the $\beta$-law in the domain $0.5 < \beta < 1.5$:
X-rays are formed in the wind, and onset radii are close to the stellar surface.
Note that a small onset radius does not
imply  that the X-ray forms \emph{only} close to the stellar surface.

\begin{figure}[htb]
\centering
  \includegraphics[width=\columnwidth]{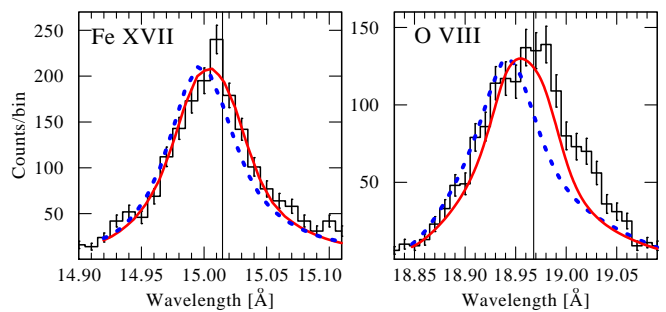}
  \caption{{\it Left panel}: Observed co-added MEG$\pm 1$ spectrum of \dori\ centered on 
Fe\,{\sc xvii} line (histogram) and two wind model lines: smooth (blue dotted line)
and clumped (red solid line). The onset 
of X-ray emitting plasma is at $1.2\,R_\ast$. {\it Right panel}: 
Same as left panel, but for the O\,{\sc viii} line}
\label{fig:clumpnoclump}
\end{figure}

While the effect of porosity is negligible at $\lambda \lesssim 10$\,\AA, 
it should generally   
be taken  into account when modeling lines at longer wavelengths, 
where the cool wind opacity increases.
Fig.\,\ref{fig:clumpnoclump} shows a comparison of the observed 
Fe\,{\sc xvii} $\lambda 15.01$\,\AA\ (left panel) and  O\,{\sc viii} 
$\lambda\lambda 18.967, 18.972$\,\AA\ (right panel)
lines with models that 
include and neglect clumping (red solid and blue dotted lines, respectively).
The wind model that 
neglects clumping is in poor agreement with the  data, 
while a better agreement is reached when clumping is assumed.  
Nevertheless, the exact profile shape is not reproduced, especially in
the case of the O\,{\sc viii} line.  In fact, there is evidence that 
the widths of the X-ray lines in \dori\ change with time (Paper II).  Our tests indicate that the observed lines
can be more accurately reproduced when fitted at each phase separately. To fit the global 
profiles, the variable line profiles would need to be averaged out. 
However, such a study, while interesting, is not the subject of this Paper.

To summarize, we illustrated that clumping generally plays a
role in shaping the observed X-ray lines, that the line profiles suggest small 
onset radii of $\sim 1.1\,R_*$, and that the X-ray emitting parcels 
are likely coupled to the cool wind.

\section{Summary and Conclusions}
\label{sec:summary}

We have performed a multi-wavelength, non-LTE spectroscopic analysis of 
the massive $\delta$~Ori~A system, which contains the visually
brightest O-type eclipsing binary in the sky and a wider tertiary component.
Our goal was to obtain accurate stellar parameters for the 
components of the system, to analyze their winds, and to 
gain a better understanding of the X-ray radiation 
emitted from the system.
Three additional studies performed within the framework of 
the current collaboration explore the X-ray properties (Paper I) and variability (Paper II)
of the system, and conduct a complete binary and
optical variabilty analysis of the system (Paper III). 

We conclude the following:

\begin{itemize}
 \item Distance-independent parameters such as $T_*$ and $\log g_*$ derived for the primary (cf.\ Table\,\ref{tab:stellarpar}) are consistent with its spectral type O9.5~II.  
        Distance-dependent parameters such as the luminosity and mass are found to be unusually low for the {\it Hipparcos} distance $d = 212$\,pc. Typical values 
       ($L_1 = 5.28\,L_\odot$, $M_1 = 24\,M_\odot$) are obtained for the distance $d = 380$\,pc of the neighboring cluster.
    These results agree well with the results reported in Paper~III, suggesting that the {\it Hipparcos} distance is strongly underestimated. 
 \item The secondary is marginally detectable in the composite spectrum. The V-band magnitude difference 	  between the primary and secondary is constrained to 
        $\Delta V_{\text{Aa1Aa2}} \sim 2\fm8$. The parameters of the secondary  suggest it is a B type dwarf. 
      The tertiary is confirmed to be a rapid rotator with $v \sin i \sim 220\,$\kms, and its parameters correspond to an early B type or a late O type subgiant. 
 \item Rapid turbulent velocities ($\sim 200$\,\kms) prevail in the wind of the primary close to the stellar surface. 
       We further find evidence for optically thick wind-inhomogeneities (``macroclumping''), affecting both strong resonance lines and X-ray lines.
 \item For a clumping factor of $D = 10$ and accounting for porosity, 
       the primary's inferred mass-loss rate is $\log \dot{M} \approx -6.4\,$\smy.
       This value provides a consistent picture for $\delta$~Ori~A along all spectral domains, from X-rays and UV to 
       the optical and radio, and is furthermore in good agreement
       with hydrodynamical predictions. 
 \item Most X-rays emerging from $\delta$ Ori A are likely intrinsic to the wind of the primary. X-ray onset radii are found to be $\sim 1.1R_*$.
\end{itemize}

While $\delta$ Ori A gives us a precious opportunity to study stellar winds and 
multiple systems of massive stars,
there are still questions left unanswered. Does the primary really have an exceptionally low mass,
luminosity, and mass-loss rate, or is the {\it Hipparcos} distance significantly
underestimated? 
Did mass-transfer occur  between the
primary Aa1 and the secondary Aa2?  Where do the clumps in the primary wind originate, 
and how can they be further quantified?
Why do we not clearly observe X-ray emission originating in the
interaction between the primary and secondary?
How did the system evolve to its current state,
and how will it continue to evolve? To answer these questions, efforts
should be made in obtaining high S/N optical and X-ray spectra which resolve the binary Aa from
the tertiary Ab, and which enable a proper disentanglement of the
system. Moreover, further studies should provide more hints regarding
the correct distance of the system.  Finally, variability studies of
P Cygni lines in the system should provide crucial, independent constraints
on the amount of inhomogeniety in the primary's wind.
Answering these questions should take us 
one step closer to a more complete understanding of stellar winds, and of the evolution 
and properties of massive 
stars in multiple systems.

\acknowledgements{We thank our anonymous referee for constructive comments which helped
to improve our paper.
TS is grateful for financial support from the Leibniz Graduate School for Quantitative 
Spectroscopy in Astrophysics, a joint project of the Leibniz Institute for Astrophysics Potsdam (AIP) 
and the institute of Physics and Astronomy of the University of Potsdam. 
LMO acknowledges support from DLR grant 50 OR 1302
We would like to thank A.\ Valeev and S.\ Fabrika for kindly providing us with an optical spectrum 
of the system.  We thank T.\ J.\ Henry and J.\ A.\ Caballero for fruitful discussions regarding 
system's distance.
MFC, JSN, and WLW are grateful for support via Chandra grant GO3-14015A and GO3-14015E.
AFJM. acknowledges financial aid from NSERC (Canada) and FRQNT (Quebec).
JMA acknowledges support from [a] the Spanish Government Ministerio de
Econom{\'\i}a y Competitividad (MINECO) through grants AYA2010-15\,081 and 
AYA2010-17\,631 and [b] the Consejer{\'\i}a de Educaci{\'o}n of the Junta de Andaluc {\'\i}a through grant
P08-TIC-4075. Caballero
NDR gratefully acknowledges his Centre du Recherche en Astrophysique du Qu\'ebec (CRAQ) fellowship.
YN acknowledges support from the Fonds National de la Recherche Scientifique (Belgium), 
the Communaut\'e Fran\c caise de Belgique, 
the PRODEX XMM and Integral contracts, and the `Action de Recherche
Concert\'ee` (CFWB-Acad\'emie Wallonie Europe).
JLH acknowledges support from NASA award NNX13AF40G and NSF award AST-0807477.
IN is supported by the Spanish Mineco under grant
AYA2012-39364-C02-01/02, and the European Union.}

\bibliography{literature}

\begin{thebibliography}{142}
\expandafter\ifx\csname natexlab\endcsname\relax\def\natexlab#1{#1}\fi

\bibitem[{{Aerts} {et~al.}(2009){Aerts}, {Puls}, {Godart}, \&
  {Dupret}}]{Aerts2009}
{Aerts}, C., {Puls}, J., {Godart}, M., \& {Dupret}, M.-A. 2009, \aap, 508, 409

\bibitem[{{Aldoretta} {et~al.}(2014){Aldoretta}, {Caballero-Nieves}, {Gies},
  {Nelan}, {Wallace}, {Hartkopf}, {Henry}, {Jao}, {Ma{\'{\i}}z Apell{\'a}niz},
  {Mason}, {Moffat}, {Norris}, {Richardson}, \& {Williams}}]{Aldoretta2014}
{Aldoretta}, E.~J., {Caballero-Nieves}, S.~M., {Gies}, D.~R., {et~al.} 2014,
  ArXiv e-prints

\bibitem[{{Anderson}(1989)}]{Anderson1989}
{Anderson}, L.~S. 1989, \apj, 339, 558

\bibitem[{{Antokhin}(2011)}]{Antokhin2011}
{Antokhin}, I.~I. 2011, Bulletin de la Societe Royale des Sciences de Liege,
  80, 549

\bibitem[{{Asplund} {et~al.}(2009){Asplund}, {Grevesse}, {Sauval}, \&
  {Scott}}]{Asplund2009}
{Asplund}, M., {Grevesse}, N., {Sauval}, A.~J., \& {Scott}, P. 2009, \araa, 47,
  481

\bibitem[{{Auger}(1923)}]{Auger1923}
{Auger}, P. 1923, C.R.A.S., 177, 169

\bibitem[{{Bagnuolo} \& {Gies}(1991)}]{Bagnuolo1991}
{Bagnuolo}, Jr., W.~G. \& {Gies}, D.~R. 1991, \apj, 376, 266

\bibitem[{{Barb{\'a}} {et~al.}(2010){Barb{\'a}}, {Gamen}, {Arias}, {Morrell},
  {Ma{\'{\i}}z Apell{\'a}niz}, {Alfaro}, {Walborn}, \& {Sota}}]{Barba2010}
{Barb{\'a}}, R.~H., {Gamen}, R., {Arias}, J.~I., {et~al.} 2010, in Revista
  Mexicana de Astronomia y Astrofisica Conference Series, Vol.~38, Revista
  Mexicana de Astronomia y Astrofisica Conference Series, 30--32

\bibitem[{{Baum} {et~al.}(1992){Baum}, {Hamann}, {Koesterke}, \&
  {Wessolowski}}]{Baum1992}
{Baum}, E., {Hamann}, W.-R., {Koesterke}, L., \& {Wessolowski}, U. 1992, \aap,
  266, 402

\bibitem[{{Blumenthal} {et~al.}(1972){Blumenthal}, {Drake}, \&
  {Tucker}}]{Blumenthal1972}
{Blumenthal}, G.~R., {Drake}, G.~W.~F., \& {Tucker}, W.~H. 1972, \apj, 172, 205

\bibitem[{{Bouret} {et~al.}(2012){Bouret}, {Hillier}, {Lanz}, \&
  {Fullerton}}]{Bouret2012}
{Bouret}, J.-C., {Hillier}, D.~J., {Lanz}, T., \& {Fullerton}, A.~W. 2012,
  \aap, 544, A67

\bibitem[{{Caballero} \& {Solano}(2008)}]{Caballero2008}
{Caballero}, J.~A. \& {Solano}, E. 2008, \aap, 485, 931

\bibitem[{{Cananzi} {et~al.}(1993){Cananzi}, {Augarde}, \&
  {Lequeux}}]{Cananzi1993}
{Cananzi}, K., {Augarde}, R., \& {Lequeux}, J. 1993, \aaps, 101, 599

\bibitem[{{Cantiello} {et~al.}(2009){Cantiello}, {Langer}, {Brott}, {de Koter},
  {Shore}, {Vink}, {Voegler}, {Lennon}, \& {Yoon}}]{Cantiello2009}
{Cantiello}, M., {Langer}, N., {Brott}, I., {et~al.} 2009, \aap, 499, 279

\bibitem[{{Cardelli} {et~al.}(1989){Cardelli}, {Clayton}, \&
  {Mathis}}]{Cardelli1989}
{Cardelli}, J.~A., {Clayton}, G.~C., \& {Mathis}, J.~S. 1989, \apj, 345, 245

\bibitem[{{Cassinelli} \& {Olson}(1979)}]{Cassinelli1979}
{Cassinelli}, J.~P. \& {Olson}, G.~L. 1979, \apj, 229, 304

\bibitem[{{Castor} {et~al.}(1975){Castor}, {Abbott}, \& {Klein}}]{CAK1975}
{Castor}, J.~I., {Abbott}, D.~C., \& {Klein}, R.~I. 1975, \apj, 195, 157

\bibitem[{{Chini} {et~al.}(2012){Chini}, {Hoffmeister}, {Nasseri}, {Stahl}, \&
  {Zinnecker}}]{Chini2012}
{Chini}, R., {Hoffmeister}, V.~H., {Nasseri}, A., {Stahl}, O., \& {Zinnecker},
  H. 2012, \mnras, 424, 1925

\bibitem[{{Cohen} {et~al.}(2014){Cohen}, {Wollman}, {Leutenegger}, {Sundqvist},
  {Fullerton}, {Zsarg{\'o}}, \& {Owocki}}]{Cohen2014}
{Cohen}, D.~H., {Wollman}, E.~E., {Leutenegger}, M.~A., {et~al.} 2014, \mnras,
  439, 908

\bibitem[{{Corcoran}(2003)}]{Corcoran2003}
{Corcoran}, M.~F. 2003, in IAU Symposium, Vol. 212, A Massive Star Odyssey:
  From Main Sequence to Supernova, ed. K.~{van der Hucht}, A.~{Herrero}, \&
  C.~{Esteban}, 130

\bibitem[{{Cutri} \& {et al.}(2012)}]{Cutri2012}
{Cutri}, R.~M. \& {et al.} 2012, VizieR Online Data Catalog, 2311, 0

\bibitem[{{de Mink} {et~al.}(2013){de Mink}, {Langer}, {Izzard}, {Sana}, \& {de
  Koter}}]{DeMink2013}
{de Mink}, S.~E., {Langer}, N., {Izzard}, R.~G., {Sana}, H., \& {de Koter}, A.
  2013, \apj, 764, 166

\bibitem[{{Eldridge} {et~al.}(2013){Eldridge}, {Fraser}, {Smartt}, {Maund}, \&
  {Crockett}}]{Eldridge2013}
{Eldridge}, J.~J., {Fraser}, M., {Smartt}, S.~J., {Maund}, J.~R., \&
  {Crockett}, R.~M. 2013, \mnras, 436, 774

\bibitem[{{ESA}(1997)}]{Hipcatalog1997}
{ESA}. 1997, VizieR Online Data Catalog, 1239, 0

\bibitem[{{Evans} {et~al.}(2004){Evans}, {Crowther}, {Fullerton}, \&
  {Hillier}}]{Evans2004}
{Evans}, C.~J., {Crowther}, P.~A., {Fullerton}, A.~W., \& {Hillier}, D.~J.
  2004, \apj, 610, 1021

\bibitem[{{Eversberg} {et~al.}(1998){Eversberg}, {Lepine}, \&
  {Moffat}}]{Eversberg1998}
{Eversberg}, T., {Lepine}, S., \& {Moffat}, A.~F.~J. 1998, \apj, 494, 799

\bibitem[{{Feldmeier} {et~al.}(2003){Feldmeier}, {Oskinova}, \&
  {Hamann}}]{Feldmeier2003}
{Feldmeier}, A., {Oskinova}, L., \& {Hamann}, W.-R. 2003, \aap, 403, 217

\bibitem[{{Feldmeier} {et~al.}(1997){Feldmeier}, {Puls}, \&
  {Pauldrach}}]{Feldmeier1997}
{Feldmeier}, A., {Puls}, J., \& {Pauldrach}, A.~W.~A. 1997, \aap, 322, 878

\bibitem[{{Friend} \& {MacGregor}(1984)}]{Friend1984}
{Friend}, D.~B. \& {MacGregor}, K.~B. 1984, \apj, 282, 591

\bibitem[{{Fullerton} {et~al.}(2006){Fullerton}, {Massa}, \&
  {Prinja}}]{Fullerton2006}
{Fullerton}, A.~W., {Massa}, D.~L., \& {Prinja}, R.~K. 2006, \apj, 637, 1025

\bibitem[{{Gabriel} \& {Jordan}(1969)}]{Gabriel1969}
{Gabriel}, A.~H. \& {Jordan}, C. 1969, \mnras, 145, 241

\bibitem[{{Gies} \& {Lambert}(1992)}]{Gies1992}
{Gies}, D.~R. \& {Lambert}, D.~L. 1992, \apj, 387, 673

\bibitem[{{Gonz{\'a}lez} \& {Levato}(2006)}]{Gonzalez2006}
{Gonz{\'a}lez}, J.~F. \& {Levato}, H. 2006, \aap, 448, 283

\bibitem[{{Gr{\"a}fener} {et~al.}(2002){Gr{\"a}fener}, {Koesterke}, \&
  {Hamann}}]{Graefener2002}
{Gr{\"a}fener}, G., {Koesterke}, L., \& {Hamann}, W.-R. 2002, \aap, 387, 244

\bibitem[{{Gray}(1975)}]{Gray1975}
{Gray}, D.~F. 1975, \apj, 202, 148

\bibitem[{{Habets} \& {Heintze}(1981)}]{Habets1981}
{Habets}, G.~M.~H.~J. \& {Heintze}, J.~R.~W. 1981, \aaps, 46, 193

\bibitem[{{Hadrava}(1995)}]{Hadrava1995}
{Hadrava}, P. 1995, \aaps, 114, 393

\bibitem[{{Hainich} {et~al.}(2014){Hainich}, {R{\"u}hling}, {Todt}, {Oskinova},
  {Liermann}, {Gr{\"a}fener}, {Foellmi}, {Schnurr}, \& {Hamann}}]{Hainich2014}
{Hainich}, R., {R{\"u}hling}, U., {Todt}, H., {et~al.} 2014, \aap, 565, A27

\bibitem[{{Hamann} \& {Gr{\"a}fener}(2004)}]{Hamann2004}
{Hamann}, W.-R. \& {Gr{\"a}fener}, G. 2004, \aap, 427, 697

\bibitem[{{Hamann} {et~al.}(2006){Hamann}, {Gr{\"a}fener}, \&
  {Liermann}}]{Hamann2006}
{Hamann}, W.-R., {Gr{\"a}fener}, G., \& {Liermann}, A. 2006, \aap, 457, 1015

\bibitem[{{Hamann} \& {Koesterke}(1998)}]{Hamann1998}
{Hamann}, W.-R. \& {Koesterke}, L. 1998, \aap, 335, 1003

\bibitem[{{Hamann} {et~al.}(1995){Hamann}, {Koesterke}, \&
  {Wessolowski}}]{Hamann1995a}
{Hamann}, W.-R., {Koesterke}, L., \& {Wessolowski}, U. 1995, \aap, 299, 151

\bibitem[{{Harvey} {et~al.}(1987){Harvey}, {Stickland}, {Howarth}, \&
  {Zuiderwijk}}]{Harvey1987}
{Harvey}, A.~S., {Stickland}, D.~J., {Howarth}, I.~D., \& {Zuiderwijk}, E.~J.
  1987, The Observatory, 107, 205

\bibitem[{{Harvin} {et~al.}(2002){Harvin}, {Gies}, {Bagnuolo}, {Penny}, \&
  {Thaller}}]{Harvin2002}
{Harvin}, J.~A., {Gies}, D.~R., {Bagnuolo}, Jr., W.~G., {Penny}, L.~R., \&
  {Thaller}, M.~L. 2002, \apj, 565, 1216

\bibitem[{{Heintz}(1980)}]{Heintz1980}
{Heintz}, W.~D. 1980, \apjs, 44, 111

\bibitem[{{Herv{\'e}} {et~al.}(2013){Herv{\'e}}, {Rauw}, \&
  {Naz{\'e}}}]{Herve2013}
{Herv{\'e}}, A., {Rauw}, G., \& {Naz{\'e}}, Y. 2013, \aap, 551, A83

\bibitem[{{Hillier}(1984)}]{Hillier1984}
{Hillier}, D.~J. 1984, \apj, 280, 744

\bibitem[{{Hillier}(1991)}]{Hillier1991_wings}
{Hillier}, D.~J. 1991, \aap, 247, 455

\bibitem[{{Hillier} {et~al.}(2012){Hillier}, {Bouret}, {Lanz}, \&
  {Busche}}]{Hillier2012}
{Hillier}, D.~J., {Bouret}, J.-C., {Lanz}, T., \& {Busche}, J.~R. 2012, \mnras,
  426, 1043

\bibitem[{{Hirschi}(2008)}]{Hirschi2008}
{Hirschi}, R. 2008, in Clumping in Hot-Star Winds, ed. W.-R. {Hamann},
  A.~{Feldmeier}, \& L.~M. {Oskinova}, 9

\bibitem[{{Horch} {et~al.}(2001){Horch}, {Ninkov}, \& {Franz}}]{Horch2001}
{Horch}, E., {Ninkov}, Z., \& {Franz}, O.~G. 2001, \aj, 121, 1583

\bibitem[{{Howarth}(1997)}]{Howarth1997}
{Howarth}, I.~D. 1997, The Observatory, 117, 335

\bibitem[{{Howarth} \& {Stevens}(2014)}]{Howarth2014}
{Howarth}, I.~D. \& {Stevens}, I.~R. 2014, \mnras, 445, 2878

\bibitem[{{Hummel} {et~al.}(2013){Hummel}, {Rivinius}, {Nieva}, {Stahl}, {van
  Belle}, \& {Zavala}}]{Hummel2013}
{Hummel}, C.~A., {Rivinius}, T., {Nieva}, M.-F., {et~al.} 2013, \aap, 554, A52

\bibitem[{{Hurley} {et~al.}(2002){Hurley}, {Tout}, \& {Pols}}]{Hurley2002}
{Hurley}, J.~R., {Tout}, C.~A., \& {Pols}, O.~R. 2002, \mnras, 329, 897

\bibitem[{{Ignace}(2001)}]{Ignace2001}
{Ignace}, R. 2001, \apjl, 549, L119

\bibitem[{{Ignace} \& {Gayley}(2002)}]{Ignace2002}
{Ignace}, R. \& {Gayley}, K.~G. 2002, \apj, 568, 954

\bibitem[{{Ignace} {et~al.}(2012){Ignace}, {Waldron}, {Cassinelli}, \&
  {Burke}}]{Ignace2012}
{Ignace}, R., {Waldron}, W.~L., {Cassinelli}, J.~P., \& {Burke}, A.~E. 2012,
  \apj, 750, 40

\bibitem[{{Koch} \& {Hrivnak}(1981)}]{Koch1987}
{Koch}, R.~H. \& {Hrivnak}, B.~J. 1981, \apj, 248, 249

\bibitem[{{Krti{\v c}ka} {et~al.}(2009){Krti{\v c}ka}, {Feldmeier}, {Oskinova},
  {Kub{\'a}t}, \& {Hamann}}]{Krticka2009}
{Krti{\v c}ka}, J., {Feldmeier}, A., {Oskinova}, L.~M., {Kub{\'a}t}, J., \&
  {Hamann}, W.-R. 2009, \aap, 508, 841

\bibitem[{{Kudritzki} {et~al.}(1989){Kudritzki}, {Pauldrach}, {Puls}, \&
  {Abbott}}]{Kudritzki1989}
{Kudritzki}, R.~P., {Pauldrach}, A., {Puls}, J., \& {Abbott}, D.~C. 1989, \aap,
  219, 205

\bibitem[{{Kudritzki} \& {Puls}(2000)}]{Kudritzki2000}
{Kudritzki}, R.-P. \& {Puls}, J. 2000, \araa, 38, 613

\bibitem[{{Lamers} {et~al.}(1999){Lamers}, {Haser}, {de Koter}, \&
  {Leitherer}}]{Lamers1999}
{Lamers}, H.~J.~G.~L.~M., {Haser}, S., {de Koter}, A., \& {Leitherer}, C. 1999,
  \apj, 516, 872

\bibitem[{{Lamers} \& {Leitherer}(1993)}]{Lamers1993}
{Lamers}, H.~J.~G.~L.~M. \& {Leitherer}, C. 1993, \apj, 412, 771

\bibitem[{{Lamers} \& {Cassinelli}(1999)}]{Cassinelli}
{Lamers}, J. G. L. M.~H. \& {Cassinelli}, J.~P. 1999, Stellar Winds (Cambridge
  University Press)

\bibitem[{{Lefever} {et~al.}(2007){Lefever}, {Puls}, \& {Aerts}}]{Lefever2007}
{Lefever}, K., {Puls}, J., \& {Aerts}, C. 2007, \aap, 463, 1093

\bibitem[{{L{\'e}pine} \& {Moffat}(1999)}]{Lepine1999}
{L{\'e}pine}, S. \& {Moffat}, A.~F.~J. 1999, \apj, 514, 909

\bibitem[{{Leutenegger} {et~al.}(2006){Leutenegger}, {Paerels}, {Kahn}, \&
  {Cohen}}]{Leutenegger2006}
{Leutenegger}, M.~A., {Paerels}, F.~B.~S., {Kahn}, S.~M., \& {Cohen}, D.~H.
  2006, \apj, 650, 1096

\bibitem[{{Lutz} \& {Kelker}(1973)}]{Lutz1973}
{Lutz}, T.~E. \& {Kelker}, D.~H. 1973, \pasp, 85, 573

\bibitem[{{Macfarlane} {et~al.}(1991){Macfarlane}, {Cassinelli}, {Welsh},
  {Vedder}, {Vallerga}, \& {Waldron}}]{Macfarlane1991}
{Macfarlane}, J.~J., {Cassinelli}, J.~P., {Welsh}, B.~Y., {et~al.} 1991, \apj,
  380, 564

\bibitem[{{Maeder}(1987)}]{Maeder1987}
{Maeder}, A. 1987, \aap, 178, 159

\bibitem[{{Maheswaran} \& {Cassinelli}(2009)}]{Maheswaran2009}
{Maheswaran}, M. \& {Cassinelli}, J.~P. 2009, \mnras, 394, 415

\bibitem[{{Ma{\'{\i}}z Apell{\'a}niz}(2010)}]{Maiz2010}
{Ma{\'{\i}}z Apell{\'a}niz}, J. 2010, \aap, 518, A1

\bibitem[{{Ma{\'{\i}}z Apell{\'a}niz} {et~al.}(2008){Ma{\'{\i}}z
  Apell{\'a}niz}, {Alfaro}, \& {Sota}}]{Maiz2008}
{Ma{\'{\i}}z Apell{\'a}niz}, J., {Alfaro}, E.~J., \& {Sota}, A. 2008, ArXiv
  e-prints

\bibitem[{{Ma{\'{\i}}z Apell{\'a}niz} {et~al.}(2014){Ma{\'{\i}}z
  Apell{\'a}niz}, {Evans}, {Barb{\'a}}, {Gr{\"a}fener}, {Bestenlehner},
  {Crowther}, {Garc{\'{\i}}a}, {Herrero}, {Sana}, {Sim{\'o}n-D{\'{\i}}az},
  {Taylor}, {van Loon}, {Vink}, \& {Walborn}}]{Maiz2014}
{Ma{\'{\i}}z Apell{\'a}niz}, J., {Evans}, C.~J., {Barb{\'a}}, R.~H., {et~al.}
  2014, \aap, 564, A63

\bibitem[{{Markova} \& {Puls}(2008)}]{Markova2008}
{Markova}, N. \& {Puls}, J. 2008, \aap, 478, 823

\bibitem[{{Markova} {et~al.}(2005){Markova}, {Puls}, {Scuderi}, \&
  {Markov}}]{Markova2005}
{Markova}, N., {Puls}, J., {Scuderi}, S., \& {Markov}, H. 2005, \aap, 440, 1133

\bibitem[{{Martins} {et~al.}(2005){Martins}, {Schaerer}, \&
  {Hillier}}]{Martins2005}
{Martins}, F., {Schaerer}, D., \& {Hillier}, D.~J. 2005, \aap, 436, 1049

\bibitem[{{Mason} {et~al.}(2009){Mason}, {Hartkopf}, {Gies}, {Henry}, \&
  {Helsel}}]{Mason2009}
{Mason}, B.~D., {Hartkopf}, W.~I., {Gies}, D.~R., {Henry}, T.~J., \& {Helsel},
  J.~W. 2009, \aj, 137, 3358

\bibitem[{{Mayer} {et~al.}(2010){Mayer}, {Harmanec}, {Wolf}, {Bo{\v z}i{\'c}},
  \& {{\v S}lechta}}]{Mayer2010}
{Mayer}, P., {Harmanec}, P., {Wolf}, M., {Bo{\v z}i{\'c}}, H., \& {{\v
  S}lechta}, M. 2010, \aap, 520, A89

\bibitem[{{McErlean} {et~al.}(1998){McErlean}, {Lennon}, \&
  {Dufton}}]{McErlean1998}
{McErlean}, N.~D., {Lennon}, D.~J., \& {Dufton}, P.~L. 1998, \aap, 329, 613

\bibitem[{{Meitner}(1922)}]{Meitner1922}
{Meitner}, L. 1922, Zeitschrift fur Physik, 9, 131

\bibitem[{{Miller} {et~al.}(2002){Miller}, {Cassinelli}, {Waldron},
  {MacFarlane}, \& {Cohen}}]{Miller2002}
{Miller}, N.~A., {Cassinelli}, J.~P., {Waldron}, W.~L., {MacFarlane}, J.~J., \&
  {Cohen}, D.~H. 2002, \apj, 577, 951

\bibitem[{{Moffat}(1994)}]{Moffat1994}
{Moffat}, A.~F.~J. 1994, \apss, 221, 467

\bibitem[{{Moffat} {et~al.}(2002){Moffat}, {Corcoran}, {Stevens}, {Skalkowski},
  {Marchenko}, {M{\"u}cke}, {Ptak}, {Koribalski}, {Brenneman}, {Mushotzky},
  {Pittard}, {Pollock}, \& {Brandner}}]{Moffat2002}
{Moffat}, A.~F.~J., {Corcoran}, M.~F., {Stevens}, I.~R., {et~al.} 2002, \apj,
  573, 191

\bibitem[{{Morel} \& {Magnenat}(1978)}]{Morel1978}
{Morel}, M. \& {Magnenat}, P. 1978, \aaps, 34, 477

\bibitem[{{Moshir} \& {et al.}(1990)}]{Moshir1990}
{Moshir}, M. \& {et al.} 1990, in IRAS Faint Source Catalogue, version 2.0
  (1990), 0

\bibitem[{{Muijres} {et~al.}(2011){Muijres}, {de Koter}, {Vink}, {Krti{\v
  c}ka}, {Kub{\'a}t}, \& {Langer}}]{Muijres2011}
{Muijres}, L.~E., {de Koter}, A., {Vink}, J.~S., {et~al.} 2011, \aap, 526, A32

\bibitem[{{Naz{\'e}}(2009)}]{Naze2009}
{Naz{\'e}}, Y. 2009, \aap, 506, 1055

\bibitem[{{Negueruela} {et~al.}(2014){Negueruela}, {Ma{\'{\i}}z Apell{\'a}niz},
  {Sim{\'o}n-D{\'{\i}}az}, ., {Alonso}, {Barb{\'a}}, ., {Mongui{\'o}},
  {Morrell}, \& nd~{Walborn}}]{Negueruela2014}
{Negueruela}, I., {Ma{\'{\i}}z Apell{\'a}niz}, J., {Sim{\'o}n-D{\'{\i}}az}, S.,
  {et~al.} 2014, ArXiv e-prints

\bibitem[{{Oskinova}(2005)}]{Oskinova2005}
{Oskinova}, L.~M. 2005, \mnras, 361, 679

\bibitem[{{Oskinova} {et~al.}(2004){Oskinova}, {Feldmeier}, \&
  {Hamann}}]{Oskinova2004}
{Oskinova}, L.~M., {Feldmeier}, A., \& {Hamann}, W.-R. 2004, \aap, 422, 675

\bibitem[{{Oskinova} {et~al.}(2011){Oskinova}, {Hamann}, {Cassinelli}, {Brown},
  \& {Todt}}]{Oskinova2011}
{Oskinova}, L.~M., {Hamann}, W.-R., {Cassinelli}, J.~P., {Brown}, J.~C., \&
  {Todt}, H. 2011, Astronomische Nachrichten, 332, 988

\bibitem[{{Oskinova} {et~al.}(2007){Oskinova}, {Hamann}, \&
  {Feldmeier}}]{Oskinova2007}
{Oskinova}, L.~M., {Hamann}, W.-R., \& {Feldmeier}, A. 2007, \aap, 476, 1331

\bibitem[{{Owocki} {et~al.}(1988){Owocki}, {Castor}, \& {Rybicki}}]{Owocki1988}
{Owocki}, S.~P., {Castor}, J.~I., \& {Rybicki}, G.~B. 1988, \apj, 335, 914

\bibitem[{{Owocki} \& {Cohen}(2006)}]{owocki2006}
{Owocki}, S.~P. \& {Cohen}, D.~H. 2006, \apj, 648, 565

\bibitem[{{Owocki} {et~al.}(2004){Owocki}, {Gayley}, \& {Shaviv}}]{Owocki2004}
{Owocki}, S.~P., {Gayley}, K.~G., \& {Shaviv}, N.~J. 2004, \apj, 616, 525

\bibitem[{{Palate} {et~al.}(2013){Palate}, {Rauw}, {Koenigsberger}, \&
  {Moreno}}]{Palate2013}
{Palate}, M., {Rauw}, G., {Koenigsberger}, G., \& {Moreno}, E. 2013, \aap, 552,
  A39

\bibitem[{{Pallavicini} {et~al.}(1981){Pallavicini}, {Golub}, {Rosner},
  {Vaiana}, {Ayres}, \& {Linsky}}]{Pallavicini1981}
{Pallavicini}, R., {Golub}, L., {Rosner}, R., {et~al.} 1981, \apj, 248, 279

\bibitem[{{Pauldrach} {et~al.}(2012){Pauldrach}, {Vanbeveren}, \&
  {Hoffmann}}]{Pauldrach2012}
{Pauldrach}, A.~W.~A., {Vanbeveren}, D., \& {Hoffmann}, T.~L. 2012, \aap, 538,
  A75

\bibitem[{{Perryman} \& {ESA}(1997)}]{Perryman1997}
{Perryman}, M.~A.~C. \& {ESA}, eds. 1997, ESA Special Publication, Vol. 1200,
  {The HIPPARCOS and TYCHO catalogues. Astrometric and photometric star
  catalogues derived from the ESA HIPPARCOS Space Astrometry Mission}

\bibitem[{{Petit} {et~al.}(2014){Petit}, {Louge}, {Th{\'e}ado}, {Paletou},
  {Manset}, {Morin}, {Marsden}, \& {Jeffers}}]{Petit2014}
{Petit}, P., {Louge}, T., {Th{\'e}ado}, S., {et~al.} 2014, \pasp, 126, 469

\bibitem[{{Pollock}(2007)}]{Pollock2007}
{Pollock}, A.~M.~T. 2007, \aap, 463, 1111

\bibitem[{{Pols} {et~al.}(1991){Pols}, {Cote}, {Waters}, \& {Heise}}]{Pols1991}
{Pols}, O.~R., {Cote}, J., {Waters}, L.~B.~F.~M., \& {Heise}, J. 1991, \aap,
  241, 419

\bibitem[{{Porquet} {et~al.}(2001){Porquet}, {Mewe}, {Dubau}, {Raassen}, \&
  {Kaastra}}]{Porquet2001}
{Porquet}, D., {Mewe}, R., {Dubau}, J., {Raassen}, A.~J.~J., \& {Kaastra},
  J.~S. 2001, \aap, 376, 1113

\bibitem[{{Prinja} \& {Massa}(2010)}]{Prinja2010}
{Prinja}, R.~K. \& {Massa}, D.~L. 2010, \aap, 521, L55

\bibitem[{{Puls}(2008)}]{Puls2008}
{Puls}, J. 2008, in IAU Symposium, Vol. 250, IAU Symposium, ed. F.~{Bresolin},
  P.~A. {Crowther}, \& J.~{Puls}, 25--38

\bibitem[{{Puls} {et~al.}(1996){Puls}, {Kudritzki}, {Herrero}, {Pauldrach},
  {Haser}, {Lennon}, {Gabler}, {Voels}, {Vilchez}, {Wachter}, \&
  {Feldmeier}}]{Puls1996}
{Puls}, J., {Kudritzki}, R.-P., {Herrero}, A., {et~al.} 1996, \aap, 305, 171

\bibitem[{{Raassen} \& {Pollock}(2013)}]{Raassen2013}
{Raassen}, A.~J.~J. \& {Pollock}, A.~M.~T. 2013, \aap, 550, A55

\bibitem[{{Raghavan} {et~al.}(2009){Raghavan}, {McAlister}, {Torres}, {Latham},
  {Mason}, {Boyajian}, {Baines}, {Williams}, {ten Brummelaar}, {Farrington},
  {Ridgway}, {Sturmann}, {Sturmann}, \& {Turner}}]{Raghavan2009}
{Raghavan}, D., {McAlister}, H.~A., {Torres}, G., {et~al.} 2009, \apj, 690, 394

\bibitem[{{Runacres} \& {Owocki}(2002)}]{Runacres2002}
{Runacres}, M.~C. \& {Owocki}, S.~P. 2002, \aap, 381, 1015

\bibitem[{{Sana} {et~al.}(2013){Sana}, {de Koter}, {de Mink}, {Dunstall},
  {Evans}, {H{\'e}nault-Brunet}, {Ma{\'{\i}}z Apell{\'a}niz},
  {Ram{\'{\i}}rez-Agudelo}, {Taylor}, {Walborn}, {Clark}, {Crowther},
  {Herrero}, {Gieles}, {Langer}, {Lennon}, \& {Vink}}]{Sana2013}
{Sana}, H., {de Koter}, A., {de Mink}, S.~E., {et~al.} 2013, \aap, 550, A107

\bibitem[{{Schmidt-Kaler}(1982)}]{Schmidt-Kaler1982}
{Schmidt-Kaler}, T. 1982, Vol. VI/2b, K. Schaifers, H.H. Voigt Eds.,
  Springer-Verlag. p. 1.

\bibitem[{{Seward} \& {Chlebowski}(1982)}]{Seward1982}
{Seward}, F.~D. \& {Chlebowski}, T. 1982, \apj, 256, 530

\bibitem[{{Shaviv}(2000)}]{Shaviv2000}
{Shaviv}, N.~J. 2000, \apjl, 532, L137

\bibitem[{{Shenar} {et~al.}(2014){Shenar}, {Hamann}, \& {Todt}}]{Shenar2014}
{Shenar}, T., {Hamann}, W.-R., \& {Todt}, H. 2014, \aap, 562, A118

\bibitem[{{Simon} \& {Sturm}(1994)}]{Simon1994}
{Simon}, K.~P. \& {Sturm}, E. 1994, \aap, 281, 286

\bibitem[{{Sim{\'o}n-D{\'{\i}}az} {et~al.}(2015){Sim{\'o}n-D{\'{\i}}az},
  {Caballero}, {Lorenzo}, {Ma{\'{\i}}z Apell{\'a}niz}, {Schneider},
  {Negueruela}, {Barb{\'a}}, {Dorda}, {Marco}, {Montes}, {Pellerin},
  {Sanchez-Bermudez}, {S{\'o}dor}, \& {Sota}}]{SimonDiaz2015}
{Sim{\'o}n-D{\'{\i}}az}, S., {Caballero}, J.~A., {Lorenzo}, J., {et~al.} 2015,
  \apj, 799, 169

\bibitem[{{Sim{\'o}n-D{\'{\i}}az} \& {Herrero}(2007)}]{SimonDiaz2007}
{Sim{\'o}n-D{\'{\i}}az}, S. \& {Herrero}, A. 2007, \aap, 468, 1063

\bibitem[{{Smartt} {et~al.}(1997){Smartt}, {Dufton}, \& {Lennon}}]{Smartt1997}
{Smartt}, S.~J., {Dufton}, P.~L., \& {Lennon}, D.~J. 1997, \aap, 326, 763

\bibitem[{{Sota} {et~al.}(2014){Sota}, {Ma{\'{\i}}z Apell{\'a}niz}, {Morrell},
  {Barb{\'a}}, {Walborn}, {Gamen}, {Arias}, \& {Alfaro}}]{Sota2014}
{Sota}, A., {Ma{\'{\i}}z Apell{\'a}niz}, J., {Morrell}, N.~I., {et~al.} 2014,
  \apjs, 211, 10

\bibitem[{{Sota} {et~al.}(2011){Sota}, {Ma{\'{\i}}z Apell{\'a}niz}, {Walborn},
  {Alfaro}, {Barb{\'a}}, {Morrell}, {Gamen}, \& {Arias}}]{Sota2011}
{Sota}, A., {Ma{\'{\i}}z Apell{\'a}niz}, J., {Walborn}, N.~R., {et~al.} 2011,
  \apjs, 193, 24

\bibitem[{{St-Louis} \& {Moffat}(2008)}]{St-Louis2008}
{St-Louis}, N. \& {Moffat}, A.~F.~J. 2008, in Clumping in Hot-Star Winds, ed.
  W.-R. {Hamann}, A.~{Feldmeier}, \& L.~M. {Oskinova}, 39

\bibitem[{{Sundqvist} \& {Owocki}(2013)}]{Sundqvist2013}
{Sundqvist}, J.~O. \& {Owocki}, S.~P. 2013, \mnras, 428, 1837

\bibitem[{{Sundqvist} {et~al.}(2011){Sundqvist}, {Puls}, {Feldmeier}, \&
  {Owocki}}]{Sundqvist2011}
{Sundqvist}, J.~O., {Puls}, J., {Feldmeier}, A., \& {Owocki}, S.~P. 2011, \aap,
  528, A64

\bibitem[{{Szabados}(1997)}]{Szabados1997}
{Szabados}, L. 1997, in ESA Special Publication, Vol. 402, Hipparcos - Venice
  '97, ed. R.~M. {Bonnet}, E.~{H{\o}g}, P.~L. {Bernacca}, L.~{Emiliani},
  A.~{Blaauw}, C.~{Turon}, J.~{Kovalevsky}, L.~{Lindegren}, H.~{Hassan},
  M.~{Bouffard}, B.~{Strim}, D.~{Heger}, M.~A.~C. {Perryman}, \& L.~{Woltjer},
  657--660

\bibitem[{{Todt} {et~al.}(2013){Todt}, {Kniazev}, {Gvaramadze}, {Hamann},
  {Buckley}, {Crause}, {Crawford}, {Gulbis}, {Hettlage}, {Hooper}, {Husser},
  {Kotze}, {Loaring}, {Nordsieck}, {O'Donoghue}, {Pickering}, {Potter},
  {Romero-Colmenero}, {Vaisanen}, {Williams}, \& {Wolf}}]{Todt2013}
{Todt}, H., {Kniazev}, A.~Y., {Gvaramadze}, V.~V., {et~al.} 2013, \mnras, 430,
  2302

\bibitem[{{Tokovinin} {et~al.}(2014){Tokovinin}, {Mason}, \&
  {Hartkopf}}]{Tokovinin2014}
{Tokovinin}, A., {Mason}, B.~D., \& {Hartkopf}, W.~I. 2014, \aj, 147, 123

\bibitem[{{Torres} {et~al.}(2011){Torres}, {Lampens}, {Fr{\'e}mat},
  {Hensberge}, {Lebreton}, \& {{\v S}koda}}]{Torres2011}
{Torres}, K.~B.~V., {Lampens}, P., {Fr{\'e}mat}, Y., {et~al.} 2011, \aap, 525,
  A50

\bibitem[{Uns\"{o}ld(1955)}]{Unsold}
Uns\"{o}ld, A. 1955, Physik der Sternatmosph\"{a}ren (Julius Springer)

\bibitem[{{{\v S}urlan} {et~al.}(2013){{\v S}urlan}, {Hamann}, {Aret},
  {Kub{\'a}t}, {Oskinova}, \& {Torres}}]{Surlan2013}
{{\v S}urlan}, B., {Hamann}, W.-R., {Aret}, A., {et~al.} 2013, \aap, 559, A130

\bibitem[{{van Altena} {et~al.}(1995){van Altena}, {Lee}, \&
  {Hoffleit}}]{Yale1995}
{van Altena}, W.~F., {Lee}, J.~T., \& {Hoffleit}, D. 1995, VizieR Online Data
  Catalog, 1174, 0

\bibitem[{{van Leeuwen}(2007)}]{VanLeeuwen2007}
{van Leeuwen}, F. 2007, \aap, 474, 653

\bibitem[{{Villamariz} \& {Herrero}(2000)}]{Villamariz2000}
{Villamariz}, M.~R. \& {Herrero}, A. 2000, \aap, 357, 597

\bibitem[{{Vink} {et~al.}(2000){Vink}, {de Koter}, \& {Lamers}}]{Vink2000}
{Vink}, J.~S., {de Koter}, A., \& {Lamers}, H.~J.~G.~L.~M. 2000, \aap, 362, 295

\bibitem[{{Walborn}(1972)}]{Walborn1972}
{Walborn}, N.~R. 1972, \aj, 77, 312

\bibitem[{{Waldron} \& {Cassinelli}(2001)}]{Waldron2001}
{Waldron}, W.~L. \& {Cassinelli}, J.~P. 2001, \apjl, 548, L45

\bibitem[{{Waldron} \& {Cassinelli}(2007)}]{Waldron2007}
{Waldron}, W.~L. \& {Cassinelli}, J.~P. 2007, \apj, 668, 456

\bibitem[{{Waldron} \& {Cassinelli}(2010)}]{Waldron2010}
{Waldron}, W.~L. \& {Cassinelli}, J.~P. 2010, \apjl, 711, L30

\bibitem[{{Weber} \& {Davis}(1967)}]{Weber1967}
{Weber}, E.~J. \& {Davis}, Jr., L. 1967, \apj, 148, 217

\bibitem[{{Williams} {et~al.}(1990){Williams}, {van der Hucht}, {Pollock},
  {Florkowski}, {van der Woerd}, \& {Wamsteker}}]{Williams1990}
{Williams}, P.~M., {van der Hucht}, K.~A., {Pollock}, A.~M.~T., {et~al.} 1990,
  \mnras, 243, 662

\bibitem[{{Zahn}(1975)}]{Zahn1975}
{Zahn}, J.-P. 1975, \aap, 41, 329

\end{thebibliography}

\end{document}